\documentclass[11pt,a4paper]{article}
\pdfoutput=1 

\usepackage[T1]{fontenc}
\usepackage[utf8]{inputenc}
\usepackage{graphicx}
\graphicspath{{./Figs/}}
\usepackage{mathtools} 
\usepackage{slashed}
\usepackage{verbatim}
\usepackage{dsfont}
\usepackage{bbold}
\usepackage{subcaption}
\usepackage{nicefrac}
\usepackage{wrapfig}
\usepackage{upgreek}
\usepackage{mathrsfs}
\usepackage{bm} 
\usepackage[sc]{mathpazo}
\usepackage{booktabs}
\usepackage{xcolor}
\usepackage{microtype} 

\usepackage{jheppub} 

\usepackage{cleveref} 
    \crefname{equation}{}{}
    \crefname{figure}{}{}

\newcommand{\evec}[1]{{\bm{#1}}} 
\newcommand{\idmat}{\mathds{1}} 

\newcommand{\overbar}[1]{\mkern 1.3mu\overline{\mkern-1.3mu#1\mkern-1.3mu}\mkern 1.3mu}

\newcommand{\im}{{\mathrm{i}}} 
\newcommand{\e}{\mathrm{e}} 
\newcommand{\sgn}{\operatorname{sgn}} 
\newcommand{\Tr}{\operatorname{Tr}} 

\newcommand{\sfrac}[2]{{\textstyle{\frac{#1}{#2}}}} 
\newcommand{\shalf}{{\textstyle \frac{1}{2}}}
\newcommand{\ihalf}{{\textstyle \frac{\im}{2}}}

\newcommand{\eg}{e.g.~}
\newcommand{\ie}{i.e.~}

\DeclareOldFontCommand{\rm}{\normalfont\rmfamily}{\mathrm}

\title{Quantum transport and the phase space structure of the Wightman functions}

\author[a,b]{Henri Jukkala,}
\author[a,b,c]{Kimmo Kainulainen}
\author[a,b]{and Olli Koskivaara}

\affiliation[a]{Department of Physics, University of Jyväskylä,\\ P.O.~Box 35 (YFL), FI-40014 Jyväskylä, Finland}
\affiliation[b]{Helsinki Institute of Physics, University of Helsinki,\\ P.O.~Box 64, FI-00014 Helsinki, Finland.}
\affiliation[c]{Theoretical Physics Department, CERN,\\ 1211 Geneva 23, Switzerland}

\emailAdd{henri.a.jukkala@student.jyu.fi}
\emailAdd{kimmo.kainulainen@jyu.fi}
\emailAdd{olli.a.koskivaara@student.jyu.fi}

\abstract{We study the phase space structure of exact quantum Wightman functions in spatially homogeneous, temporally varying systems. In addition to the usual mass shells, the Wightman functions display additional coherence shells around zero frequency $k_0 = 0$, which carry the information of the local quantum coherence of particle-antiparticle pairs. We find also other structures, which encode non-local correlations in time, and discuss their role and decoherence. We give a simple derivation of the cQPA formalism, a set of quantum transport equations, that can be used to study interacting systems including the local quantum coherence. We compute quantum currents created by a temporal change in a particle's mass, comparing the exact Wightman function approach, the cQPA and the semiclassical methods. We find that the semiclassical approximation, which is fully encompassed by the cQPA, works surprisingly well even for very sharp temporal features. This is encouraging for the application of semiclassical methods in electroweak baryogenesis with strong phase transitions.}


\preprint{CERN-TH-2019-173}
\arxivnumber{1910.10979}

\begin{document}
\maketitle

%
\section{Introduction}
\label{sec:intro}
%

Quantum coherence plays an important role in many physical problems in cosmology. Examples include CP-violating particle-wall interactions during the electroweak phase transition, out-of-equilibrium decay of nearly degenerate heavy neutrinos during leptogenesis, particle production during phase transitions and reheating at the end of inflation. The key quantity in the analysis of such intrinsically quantum systems is the two-point correlation function, whose evolution is described by the Schwinger--Dyson equations~\cite{Schwinger:1960qe,Keldysh:1964ud}, or in the phase space picture by the Kadanoff--Baym equations~\cite{Baym:1961zz,Mahan,Prokopec:2003pj}. The phase space picture in particular has provided a useful basis for deriving approximate transport formalisms, the prime example being the standard Boltzmann theory.

In this paper we study an exact, damped, spatially homogeneous and isotropic two-point correlation function of a fermion with a possibly complex, time-varying mass term. We show that the mixed representation correlation function contains novel shell structures which carry information about different types of quantum coherences. For example we find a shell at $k_0=0$, which encodes the information of a coherently mixing particle-antiparticle system. This shell was previously seen in the context of the coherent quasiparticle approximation (cQPA)~\cite{Herranen:2008hi,Herranen:2008hu,Herranen:2008di,Herranen:2009zi,Herranen:2009xi,Herranen:2010mh,Fidler:2011yq} in the spectral limit, but our derivation is more general, being exact in the non-interacting case. In addition we find also other shell-structures, corresponding to non-local (in the relative time coordinate), long range correlations.

All phase space structures depend sensitively on the existence and the magnitude of damping. In the non-interacting case non-local coherences dominate the system, preventing a free particle interpretation of the phase space structure in non-trivial backgrounds. Damping suppresses the non-local coherences and leads to the emergence of a local limit for time intervals $\Delta t > 1/\Gamma$, where $\Gamma$ is the damping width. For small enough $\Gamma$ the local correlation function can be well approximated by a spectral ansatz, leading to the cQPA-picture mentioned above.

We will introduce a new, elegant way to reorganise the gradient expansion in the mixed representation Kadanoff--Baym equations. We then use it to give a simple derivation of the cQPA equations complete with explicit collision integrals for arbitrary types of interactions. These equations are one of the main results of this paper: they generalise the usual Boltzmann transport theory to systems including coherent particle-antiparticle states. In particular we argue that the cQPA completely encompasses the well known semiclassical effects. Possible applications of these equations include baryogenesis during phase transitions and particle production during and after inflation.

We compute the axial current densities using the exact mixed representation correlation functions as well as their cQPA counterparts and compare these to the ones obtained in the semiclassical approximation. We find that the semiclassical methods work reasonably well even in systems where the relevant modes have a wavelength as small as half of the wall width.\footnote{Throughout this paper we use the word `wall' to refer to the temporal transition in the mass, see e.g. figure~\ref{fig:tanh1}.} This is encouraging for the application of semiclassical methods in the related problem of electroweak phase baryogenesis with very strong electroweak phase transitions. These typically create sharp transition walls and are often encountered in the context of models producing large, observable gravitational wave signals~\cite{Hindmarsh:2013xza,Hindmarsh:2015qta,Hindmarsh:2017gnf,Cutting:2018tjt,Dorsch:2016nrg,Vaskonen:2016yiu}. 

This paper is organised as follows: in section~\ref{sec:Wfunction} we first review the derivation of the cQPA formalism including the spectral Wightman functions. In section~\ref{sec:qmwigner} we construct the exact free Wightman function from mode functions, generalised to account for the damping. Some numerical examples for the phase space solutions are shown in section~\ref{sec:phasespace}. In section~\ref{sec:currents}, we compute and compare currents in different approximations in the non-interacting case. In section~\ref{sec:cQPA_with_collisions} we present cQPA transport equations in the interacting case with explicit expressions for collision terms and compute cQPA currents with interactions. Finally, in section~\ref{sec:conclusions}, we give our conclusions.

%
\section{Wightman functions and cQPA}
\label{sec:Wfunction}
%

We are using the Schwinger--Keldysh formalism~\cite{Schwinger:1960qe,Keldysh:1964ud} of finite temperature field theory. The key quantities are the two-point Wightman functions
\begin{equation}
\begin{split}
\im S^<(u,v) &= \bigl\langle\overbar{\psi}(v)\psi(u)\bigr\rangle \text{,} \\
\im S^>(u,v) &= \bigl\langle\psi(u)\overline{\psi}(v)\bigr\rangle \text{,}
\end{split}
\end{equation}
which describe the quantum statistical properties of the non-equilibrium system.\footnote{Note that we define the Wightman function $S^<$ with a positive sign. We also suppress Dirac indices when there is no danger of confusion.} We also need the retarded and advanced correlation functions $\im S^{\mathrm{r}}(u,v) = 2\theta(u_0 - v_0)\mathcal{A}(u,v)$ and $\im S^{\mathrm{a}}(u,v)= - 2\theta(v_0 - u_0)\mathcal{A}(u,v)$, where the \emph{spectral function} is $\mathcal{A} \equiv \shalf\langle\{ \psi(u),\overbar{\psi}(v)\}\rangle = \ihalf(S^> + S^<) = \ihalf(S^{\mathrm{r}} - S^{\mathrm{a}})$.

To get a phase space description of the system we perform the Wigner transformation
\begin{equation}
\label{eq:wignertot}
    S(k,x) \equiv \int\mathrm{d}^4r\,\e^{\im k \cdot r}S\bigl(
        x + \sfrac{r}{2}, x - \sfrac{r}{2}
    \bigr) \text{,}
\end{equation}
where $r\equiv u-v$ and $x \equiv \frac{1}{2}(u+v)$ are the relative and average coordinates, corresponding to microscopic and macroscopic scales, respectively. In this mixed Wigner representation correlation functions obey the Kadanoff--Baym equations~\cite{Baym:1961zz}
\begin{alignat}{3}
   \bigl(\slashed k + \ihalf \slashed\partial\bigr) & S^{p}
   &&{}- \e^{-\im\Diamond} \{ \Sigma^{p} \}\{ S^{p} \} &&= 1
   \text{,} \label{eq:pole-wigner} \\
   \bigl(\slashed k + \ihalf \slashed\partial\bigr) & S^{s}
   &&{}- \e^{-\im\Diamond} \{ \Sigma^{\mathrm{r}} \}\{ S^{s} \} &&=
      \e^{-\im\Diamond} \{ \Sigma ^{s} \}\{ S^{\mathrm{a}} \}
   \text{,} \label{eq:kb-wigner}
\end{alignat}
where $s={<},{>}$ and $p=\mathrm{r},\mathrm{a}$ refer to the retarded and advanced functions, respectively, $\Sigma$ is the fermion self-energy and $\Diamond\{f\}\{g\} \equiv \frac{1}{2}[\partial_x f \cdot \partial_k g -\partial_k f \cdot \partial_x g]$ is the Moyal product. Note that we absorb the mass terms into the singular parts of $\Sigma^{\rm r,a}$, unless explicitly stated otherwise.

Moyal products are not the optimal way for organising the gradient expansions, and we find it useful to introduce another self-energy function:
\begin{equation}
    \Sigma_{\rm out}(k,x) \equiv \int{\rm d}^4z\,\e^{\im k\cdot(x-z)} \Sigma(x,z)
    = \e^{\ihalf\partial_x^\Sigma \cdot \partial_k^\Sigma} \Sigma(k,x)
    \text{.} 
\label{eq:SigmaOut}
\end{equation}
Using equation~\eqref{eq:SigmaOut} we can rewrite Moyal products in a form that reorganises the gradients into total $k$-derivatives controlled by the scale of variation of $\Sigma$, while all dependence on (dynamical) gradients acting on $S$ is fully accounted for by iterative resummation:
\begin{alignat}{3}
   \slashed {\hat K} & S^{p}
   &&{}- \e^{-\ihalf\partial_x^\Sigma \cdot \partial_k}[\Sigma^{p}_{\rm out}({\hat K},x) S^p]
   &&= 1
   \text{,} \label{PoleEqMix2} \\
   \slashed {\hat K} & S^{s}
   &&{}- \e^{-\ihalf\partial_x^\Sigma \cdot \partial_k}[\Sigma^{\mathrm{r}}_{\rm out} ({\hat K},x) S^{s} ]
    &&= \e^{-\ihalf\partial_x^\Sigma \cdot \partial_k}[\Sigma^{s}_{\rm out} ({\hat K},x) S^{\mathrm{a}}]
   \text{,} \label{StatEqMix2}
\end{alignat}
where ${\hat K} \equiv k + \ihalf \partial_x$. This form of the Kadanoff--Baym equations is particularly well suited for obtaining finite order expansions and iterative solutions. The mass operator is included in the singular part $\Sigma_{\rm sg}$ of the retarded/advanced self-energy functions: 
\begin{equation}
\Sigma^{\rm r,a}(k,x) = \Sigma_{\rm sg}(x) + \Sigma^{\rm H}_{\rm nsg}(k,x) \mp \im\Sigma^{\mathcal{A}}(k,x)
\text{,}\
\end{equation}
where $\Sigma^{\rm H}_{\rm nsg}$ is the non-singular Hermitian part and $\Sigma^{\mathcal{A}}$ is the anti-Hermitian part of the self-energy. To be specific, we consider a fermion field with a complex, spacetime-dependent mass $m(x)$:
\begin{equation}
    \mathcal{L} = \im\overbar{\psi}\slashed{\partial}\psi
    - m^*(x)\overbar{\psi}_{\rm R}\psi_{\rm L} - m(x)\overbar{\psi}_{\rm L}\psi_{\rm R}
    \text{,}
    \label{eq:Lagrangian}
\end{equation}
where $\psi_{\mathrm{L},\mathrm{R}} \equiv \frac{1}{2}(1\mp\gamma^5)\psi$. In the Wigner representation the spacetime-dependent mass gives rise to an operator:
\begin{equation}
({\hat m}_{\rm R} + \im \gamma^5 {\hat m}_{\rm I}) S(k,x) \equiv \e^{-\ihalf\partial_x^m \cdot \partial_k}\left\{[ m_{\rm R}(x) + \im\gamma^5m_{\rm I}(x)] S(k,x)\right\} \text{,}
\label{eq:massoperator}
\end{equation}
where $m_{\rm R}(x)$ and $m_{\rm I}(x)$ are the real and imaginary parts of $m(x)$, respectively.

Equations~\cref{PoleEqMix2,StatEqMix2} are practically impossible to solve exactly and one needs to find approximation schemes that maintain the essential physics at hand. The cQPA developed in refs.~\cite{Herranen:2008hu,Herranen:2008di,Herranen:2009zi,Herranen:2009xi,Herranen:2010mh,Fidler:2011yq,Herranen:2008hi} is one such scheme, which allows to study particular non-equilibrium systems with quantum coherence. The crux of the cQPA is to solve equations~\cref{PoleEqMix2,StatEqMix2} in two steps. First one solves for the phase space structure of the system at the lowest order in gradients and ignoring collision terms. This leads to spectral solutions for both pole and Wightman functions, where the latter contain new coherence shells in addition to the usual mass shell solutions. In the second step, one inserts these solutions back to the full equations, which are then reduced to a set of Boltzmann-like equations for generalised particle distribution functions~\cite{Herranen:2009xi,Herranen:2010mh}.

%
\subsection{cQPA-solution in a spatially homogeneous system}
\label{sec:cQPA}
%

Let us consider a spatially homogeneous and isotropic system, where $m(x) \rightarrow m(t)$ in equations~\cref{eq:Lagrangian,eq:massoperator}. The Wigner transform~\eqref{eq:wignertot} with respect to spatial coordinates then reduces to a Fourier transform, and we will denote the Wigner transform $S(k,x)$ as $S_{\evec{k}}(k_0,t)$. We also consider explicitly only the equation for $S^<$, as the derivation for $S^>$ is completely analogous. At first we will ignore interactions and work to the lowest order in gradients. The Hermitian part of equation~\eqref{StatEqMix2} for $\overbar S^<\equiv \im S^<\gamma^0$ then reduces to
\begin{equation}
    2 k_0 \overbar S_\evec{k}^<(k_0,t) = \{H_\evec{k}(t),\overbar S_\evec{k}^<(k_0,t)\}
    \text{,} 
    \label{eq:free_equations1}
\end{equation}
where $H_\evec{k}(t) \equiv \evec{\alpha}\cdot\evec{k} + \gamma^0[m_{\rm R}(t) + \im \gamma^5 m_{\rm I}(t)]$ is the free Dirac Hamiltonian.

\begin{figure}[t!]
\begin{center}
\includegraphics[scale=0.95]{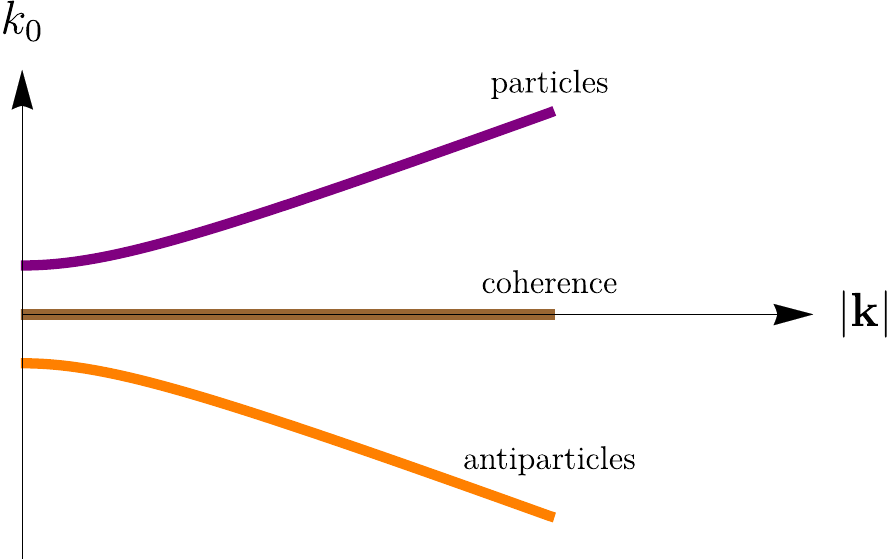}
\end{center}
\caption{The shell structure of the cQPA Wightman function $S_\evec{k}^<(k_0,t)$, showing the particle shell where $k_0=\omega_{\evec{k}}$ (purple), the antiparticle shell where $k_0=-\omega_{\evec{k}}$ (orange) and the particle-antiparticle coherence shell where $k_0=0$ (brown).}
\label{fig:cqpastr}
\end{figure}

In spatially homogeneous and isotropic systems the Wightman functions have 8 independent components and can be parametrised without any loss of generality as follows:
\begin{equation}
    \overbar S_\evec{k}^<(k_0,t) \equiv
    \sum_{h,\pm,\pm'} P^{\scriptscriptstyle (4)}_{h\evec{k}}
    P^\pm_\evec{k} \gamma^0 P^{\pm'}_\evec{k} \, D^{\pm\pm'}_{h\evec{k}}(k_0,t) \text{,}
   \label{eq:wightman-projection-parametrisation}
\end{equation}
where the helicity and energy projection operators are defined, respectively, as
\begin{equation}
   P^{\scriptscriptstyle (4)}_{h\evec{k}} \equiv
   \frac{1}{2}\bigl(\idmat + h \evec{\alpha} \cdot \hat{\evec{k}}\gamma^5\bigr) \text{,}
   \qquad
   P^\pm_{\evec{k}} \equiv
   \frac{1}{2}\biggl(\idmat \pm \frac{H_\evec{k}}{\omega_\evec{k}}\biggr)\text{,}
\label{eq:energy-projection-operator}
\end{equation}
with $\hat{\evec{k}} \equiv \evec{k}/|\evec{k}|$ and $\omega_\evec{k} \equiv \sqrt{\evec{k}^2 + |m(t)|^2}$. Inserting the parametrisation~\eqref{eq:wightman-projection-parametrisation} to equation~\eqref{eq:free_equations1} gives algebraic constraints to the time- and energy-dependent coefficient functions $D^{\pm\pm'}_{h\evec{k}}(k_0,t)$:
\begin{equation}
\begin{aligned}
   (k_0 \mp \omega_\evec{k}) D^{\pm\pm}_{h\evec{k}}(k_0,t) &= 0 \text{,} \\
   k_0 D^{\pm\mp}_{h\evec{k}}(k_0,t) &= 0 \text{.}
\end{aligned}
\end{equation}
Functions $D^{\pm\pm}_{h\evec{k}}(k_0,t) \propto \delta(k_0\mp\omega_\evec{k})$ correspond to the usual mass shell excitations, while $D^{\pm\mp}_{h\evec{k}}(k_0,t) \propto \delta(k_0)$ are the new coherence functions found in refs.~\cite{Herranen:2008hu,Herranen:2008di,Herranen:2009zi,Herranen:2009xi,Herranen:2010mh}. The spectral cQPA-solution can then be written as:
\begin{equation}
    {\overbar S}_\evec{k}^<(k_0,t) =
    2\uppi \sum_{h,\pm} \left[
        P^{m\pm}_{h\evec{k}}
        f^{m\pm}_{h\evec{k}} \delta(k_0 \mp \omega_\evec{k})
        + P^{c\pm}_{h\evec{k}} f^{c\pm}_{h\evec{k}} \delta(k_0)
    \right] \text{.}
    \label{eq:cQPA-correlator-free-particle}
\end{equation}
where we defined the projection operators 
\begin{equation}
\begin{aligned}
P^{m\pm}_{h\evec{k}} &\equiv \pm \frac{\omega_\evec{k}}{m_{\rm R}} P^{\scriptscriptstyle (4)}_{h\evec{k}} P^\pm_\evec{k} \gamma^0 P^\pm_\evec{k} = P^{\scriptscriptstyle (4)}_{h\evec{k}}P^\pm_\evec{k} \text{,}
\\
P^{c\pm}_{h\evec{k}} &\equiv P^{\scriptscriptstyle (4)}_{h\evec{k}} P^\pm_\evec{k} \gamma^0 P^\mp_\evec{k}
= P^{\scriptscriptstyle (4)}_{h\evec{k}}\left(\gamma^0 \pm \frac{m_{\rm R}}{\omega_\evec{k}} \right)P^\mp_\evec{k} \text{.} 
\end{aligned}
\end{equation}
With this normalisation the mass shell functions $f^{m\pm}_{h\evec{k}}(t)$ coincide with the usual Fermi--Dirac distributions in the thermal limit: $f^{m\pm}_{h\evec{k}} \to f_{\rm eq}(\pm\omega_{\evec{k}})$, where $f_{\rm eq}(k_0)\equiv(\e^{k_0/T} + 1)^{-1}$. Note that due to the Hermiticity of $\overbar S_{h\evec{k}}^<(k_0,t)$ the shell functions obey $(f_{h\evec{k}}^{m\pm})^* =f_{h\evec{k}}^{m\pm}$ and $(f_{h\evec{k}}^{c\pm})^* =f_{h\evec{k}}^{c\mp}$. The phase space structure of the cQPA Wightman functions is shown in figure~\ref{fig:cqpastr}.

%
%
%
%

The cQPA evolution equations are then obtained by inserting the spectral ansatz~\eqref{eq:cQPA-correlator-free-particle} to the anti-Hermitian part of equation~\eqref{StatEqMix2}, now including all gradients and interaction terms, and integrating over the energy. However, let us again first consider this equation in the non-interacting limit and to lowest order in gradients:
\begin{equation}
    \im \partial_t {\overbar S}_{\evec{k}}^<(k_0,t) = \bigl[
        H_\evec{k}, {\overbar S}_{\evec{k}}^<(k_0,t)
    \bigr] \text{.}
\label{eq:free_equations2}
\end{equation}
Substituting the spectral solution~\eqref{eq:cQPA-correlator-free-particle} for ${\overbar S}_{\evec{k}}^<(k_0,t)$ to equation~\eqref{eq:free_equations2} and integrating over $k_0$ it is easy to derive the leading behaviour of the shell-functions:
\begin{equation}
\begin{aligned}
    \partial_t f^{m\pm}_{h\evec{k}} &= \ldots \text{,} \\
    \partial_t f^{c\pm}_{h\evec{k}} &= \mp 2\im\omega_\evec{k} 
    f^{c\pm}_{h\evec{k}} + \ldots \text{,}
    \label{eq:free_equations3}
\end{aligned}
\end{equation}
where the ellipses denote terms proportional to gradient terms (and eventually self-energy terms when interactions are included).

The point we wish to make here is that the coherence shell solutions $ f^{c\pm}_{h\evec{k}}$ are oscillating rapidly with frequencies that are not suppressed by gradients. Anticipating this oscillation was the reason for our careful organisation of gradient terms in equations~\cref{PoleEqMix2,StatEqMix2}: whenever the operator $\hat K_0 = k_0 + \frac{\im}{2}\partial_t$ is acting on a coherence shell function $f^{c\pm}_{h\evec{k}}$, one must replace $\hat K_0 \rightarrow k_0 \pm \omega_\evec{k}$ as the effective momentum argument of the operator, at the \emph{lowest} order in gradients. Indeed, in cQPA:
\begin{align}
& \int \frac{{\rm d}k_0}{2\uppi}\, {\e^{-\ihalf\partial_x^\Sigma \cdot \partial_k}} \Bigl[
    \Sigma_{{\rm out},\evec{k}}\bigl(k_0 + \ihalf\partial_t\bigr) {\overbar S}^<_{h\evec{k}}(k_0,t)
\Bigr]
\notag \\
&= \sum_{\pm} \int {\rm d}k_0 \, \Sigma_{{\rm out},\evec{k}}\bigl(k_0 + \ihalf\partial_t\bigr) \Bigl[
    P^{m\pm}_{h\evec{k}}(t) f^{m\pm}_{h\evec{k}}(t) \delta(k_0 \mp \omega_{\evec{k}})
    + P^{c\pm}_{h\evec{k}}(t)f^{c\pm}_{h\evec{k}}(t)\delta(k_0)
\Bigr]
\notag \\
&\simeq \sum_{\pm} \int {\rm d}k_0 \Bigl[
    \Sigma_{{\rm out},\evec{k}}(k_0) P^{m\pm}_{h\evec{k}}(t)f^{m\pm}_{h\evec{k}}(t)\delta(k_0 \mp \omega_{\evec{k}})
    +\Sigma_{{\rm out},\evec{k}}(k_0 \pm \omega_{\evec{k}}) P^{c\pm}_{h\evec{k}}(t)f^{c\pm}_{h\evec{k}}(t)\delta(k_0)
\Bigr]
\notag \\
&= \sum_{\pm}\, \Sigma_{{\rm out},\evec{k}}(\pm \omega_{\evec{k}}) \Bigl[
    P^{m\pm}_{h\evec{k}}(t)f^{m\pm}_{h\evec{k}}(t)
    + P^{c\pm}_{h\evec{k}}(t)f^{c\pm}_{h\evec{k}}(t)
\Bigr]
\label{eq:inter_equations1}
\end{align}
for a generic self-energy function $\Sigma$. That is, coherence shell projections are not evaluated at the shell $k_0=0$, but on the mass shells instead. It would be straightforward to include higher order gradient corrections to shell positions generated by the $\hat K_0$-operator, but doing so consistently, we should also solve the cQPA-ansatz to higher order in gradients. The gradient corrections to collision terms arising from such an expansion (collisional source terms) were studied in ref.~\cite{Prokopec:2004ic} for the electroweak baryogenesis problem using semiclassical methods. They were in general found to be very small and we shall not pursue them here further. For the same reason we shall, in what follows, set $\Sigma_{{\rm out},\evec{k}} \rightarrow \Sigma_{\evec{k}}$, dropping the corrections coming from the expansion of the $\Sigma_{\rm out}$-function in equation~\eqref{eq:SigmaOut}.\footnote{Note however that the expansion of $\Sigma_{{\rm out},\evec{k}}$ may contain lowest order gradients that need to be resummed in the same way as we did above in equation~\eqref{eq:inter_equations1}. This is the case whenever the self-energy function contains an internal propagator containing the coherence function connected to the external leg in the diagram. For more details see refs.~\cite{Herranen:2010mh,Fidler:2011yq}.}

We will also work with the vacuum dispersion relations, setting $\Sigma_{\rm nsg}^{\rm H}\rightarrow 0$ and $\Sigma_{\rm sg}\rightarrow m_{\rm R} + \im\gamma^5 m_{\rm I}$. Furthermore, we shall drop the term $\propto S_{\rm H} \Sigma^<$, as this is required by the consistency of the spectral limit with respect to the pole equations~\cite{Herranen:2008hu}. With these simplifications it is now straightforward to show that the full cQPA equations can be written as
\begin{subequations}
\label{eq:cqpa_singleflav1}
\begin{align}
\label{eq:cqpa_singleflav1_a}
\partial_t f^{m\pm}_{h\evec{k}} ={}& \pm\frac{1}{2}\sum_s \dot{\Phi}_{h\evec{k}}^s f^{cs}_{h\evec{k}} + \Tr\bigl[{\cal C}_{\rm coll}P^{m\pm}_{h\evec{k}}\bigr] \text{,} \\
\label{eq:cqpa_singleflav1_b}
\partial_t f_{h\evec{k}}^{c\pm} ={}& \mp 2\im\omega_{\evec{k}} f_{h\evec{k}}^{c\pm} + \xi_{\evec{k}} \dot{\Phi}_{h\evec{k}}^{\mp} \left[\frac{m_{\mathrm{R}}}{\omega_{\evec{k}}} f_{h\evec{k}}^{c\pm} - \frac{1}{2}\left(f_{h\evec{k}}^{m+} - f_{h\evec{k}}^{m-}\right) \right]
+ \xi_{\evec{k}}\Tr\bigl[{\cal C}_{\rm coll}P^{c\mp}_{h\evec{k}}\bigr] \text{,}
\end{align}
\end{subequations}
where
\begin{equation}
{\cal C}_{\rm coll} = \sum_{h,s} \Bigl[ \Bigl(\frac{1}{2}\overbar\Sigma^<_{\evec{k}}(s\omega_{\evec{k}})
    - f^{ms}_{h\evec{k}}\,\overbar\Sigma^{\cal A}_{\evec{k}}(s\omega_{\evec{k}})\Bigr) P^{ms}_{h\evec{k}}
    - f^{cs}_{h\evec{k}}\,\overbar\Sigma^{\cal A}_{\evec{k}}(s\omega_{\evec{k}}) P^{cs}_{h\evec{k}}\Bigr]
    + {\rm h.c.}
\end{equation}
and we defined 
\begin{equation}
    \dot{\Phi}_{h\evec{k}}^{\pm} \equiv \partial_t \biggl(\frac{m_{\mathrm{R}}}{\omega_{\evec{k}}}\biggr)
        \pm \im \frac{h|\evec{k}|}{\omega_{\evec{k}}^2}\, \partial_t m_{\mathrm{I}} \text{,}\hspace{3em}
    \xi_{\evec{k}} \equiv \frac{\omega_{\evec{k}}^2}{\omega_{\evec{k}}^2 - m_{\mathrm{R}}^2} \text{.}
\label{eq:phidots}
\end{equation}
We shall return to study interacting theories in section~\ref{sec:cQPA_with_collisions}. For now, we shall take a closer look into the phase space structure of the exact non-interacting Wightman functions.

%
\section{Constructing the exact Wightman function}
\label{sec:qmwigner}
%

In the previous section we showed that Wightman functions may acquire novel  phase space structures in the spectral limit. The new coherence functions $f^{c\pm}_{h\evec{k}}$ on the $k_0=0$ shell describe quantum coherence in correlated particle-antiparticle states. These correlations can be interpreted in terms of squeezed states and the functions $f^{c\pm}_{h\evec{k}}$ can be related to Bogolyubov coefficients~\cite{Fidler:2011yq}. Condensation of the coherence information onto a sharp phase space shell is still surprising. It is therefore of interest to see how such structures arise in an exactly solvable system.

%
\subsection{Non-interacting Wightman function}
%

The Lagrangian density~\eqref{eq:Lagrangian} provides a suitable system for our study. In the spatially homogeneous case it implies the equation of motion
\begin{equation}
\im\slashed{\partial} \psi - m^*(t) \psi_{\mathrm{L}} - m(t)\psi_{\mathrm{R}} = 0 \text{.}
\label{eq:DiracEquation}
\end{equation}
We quantise this model with the usual canonical procedure.  Because three-momentum $\evec{k}$ and helicity $h$ are conserved, the field operator $\hat\psi(x)$ may be expanded in terms of mode functions as
\begin{equation}
    \hat\psi_{\rm free}(t,\evec{x})
    = \sum_{h} \int\frac{\mathrm{d}^3\evec{k}}{(2\uppi)^32\omega_{-}}\Bigl[
        \hat{a}_{h\evec{k}} U_{h\evec{k}}(t)\e^{\im\evec{k}\cdot\evec{x}}
        + \hat{b}^\dagger_{h\evec{k}} V_{h\evec{k}}(t)\e^{-\im\evec{k}\cdot\evec{x}}
    \Bigr] \text{,}
\label{eq:FreeFieldOperator}
\end{equation}
where $\omega_{-} = \sqrt{\evec{k}^2 + |m(-\infty)|^2}$. The vacuum state is annihilated as $\hat a_{h\evec{k}}|\Omega\rangle = \hat b_{h\evec{k}}|\Omega\rangle = 0$ and our normalisation is such that
\begin{equation}
\begin{split}
\{\hat a_{h\evec{k}},\hat a^\dagger_{h'{\evec{k}}^\prime}\}= (2\uppi)^3 2\omega_{-}\delta^{(3)}({\evec{k}}- {\evec{k}}^\prime) \delta_{hh^\prime} \text{,} \\
\{\hat b_{h\evec{k}},\hat b^\dagger_{h'{\evec{k}}^\prime}\}=  (2\uppi)^3 2\omega_{-}\delta^{(3)}({\evec{k}}-{\evec{k}}^\prime)\delta_{hh^\prime} \text{,}
\end{split}
\end{equation}
while all other anticommutators vanish. The normalisation of the spinor \(\hat\psi_{\rm free}\) is chosen to be such that
\begin{equation}
\{\hat\psi_{{\rm free}, \alpha}(t,\evec{x}),\hat\psi_{{\rm free}, \beta}^\dagger(t,\evec{y}) \}
= \delta_{\alpha\beta} \delta^{(3)}(\evec{x} - \evec{y}) \text{,}
\end{equation}
with the mode functions \(U_{h\evec{k}}\) and \(V_{h\evec{k}}\) normalised accordingly. The particle and antiparticle spinors can be decomposed in terms of helicity as follows:
\begin{equation}
U_{h\evec{k}}(t) = 
 \begin{bmatrix}
   \eta_{h\evec{k}}(t) \\
   \zeta_{h\evec{k}}(t)
 \end{bmatrix}
\otimes \xi_{h\evec{k}}\text{,} \qquad
V_{h\evec{k}}(t) = 
 \begin{bmatrix}
   \overbar{\eta}_{h\evec{k}}(t) \\  
   \overbar{\zeta}_{h\evec{k}}(t)
\end{bmatrix}
\otimes \xi_{h\evec{k}}\text{,}
\label{eq:decompose}
\end{equation}
where $\xi_{h\evec{k}}$ are the eigenfunctions of helicity satisfying
\begin{equation}
(\evec{\sigma} \cdot \hat{\evec{k}}) \xi_{h\evec{k}} = h\,\xi_{h\evec{k}}\text{,} \hspace{3em} h=\pm1\text{,}
\end{equation}
and $\eta_{h\evec{k}}$, $\zeta_{h\evec{k}}$, $\overbar{\eta}_{h\evec{k}}$ and $\overbar{\zeta}_{h\evec{k}}$ are yet unknown mode functions that depend on $m(t)$.\footnote{We are using the chiral basis, where the Dirac matrices are given by $\gamma^0 = \rho^1 \otimes \idmat$, $\gamma^i = \im \rho^2 \otimes \sigma^i$ and $\gamma^5  = -\rho^3 \otimes \idmat$. Here both $\rho^i$ and $\sigma^i$ are just the usual $2\times 2$ Pauli matrices. The former encode the chiral and the latter the helicity degrees of freedom of a given spinor.} The particle mode functions $\eta_{h\evec{k}}$ and $\zeta_{h\evec{k}}$ satisfy the equations
\begin{subequations}
\label{eq:ChiralComponentEom}
\begin{align}
\im\partial_t\eta_{h\evec{k}} + h|\evec{k}|\eta_{h\evec{k}} &= m(t)\zeta_{h\evec{k}}\text{,} \\
\im\partial_t\zeta_{h\evec{k}} - h|\evec{k}|\zeta_{h\evec{k}} &= m^*(t)\eta_{h\evec{k}}\text{,}
\end{align}
\end{subequations}
while the equations for the antiparticle mode functions $\overbar{\eta}_{h\evec{k}}$ and $\overbar{\zeta}_{h\evec{k}}$ contained in $V_{h\evec{k}}(t)$ can be obtained from equations~\eqref{eq:ChiralComponentEom} by the replacements $h\rightarrow -h$ and $m \rightarrow -m^*$.

The exact Wightman functions for the non-interacting system can now be constructed as expectation values of field operators in the vacuum defined by our annihilation operators. While both Wightman functions $S^>$ and $S^<$ contain the same degrees of freedom, the positive energy solutions, which we shall be using as an example below, are most straightforward to identify from $S^>$. Continuing to work in the helicity basis we find
\begin{equation}
\label{eq:UndampedWightmanFunction1}
    \im S_{hh'\evec{k}}^>(k_0,t)
    = \int{\rm d}^4r\,\e^{\im k_0 r_0 - \im\evec{k}\cdot\evec{r}} \bigl\langle\Omega\bigl|
        \hat{\psi}_{h,\mathrm{free}}\bigl(x+\sfrac{r}{2}\bigr) \hat{\overbar\psi}_{h^\prime,\mathrm{free}} \bigl(x-\sfrac{r}{2}\bigr)
    \bigr|\Omega\bigr\rangle \text{.}
\end{equation}
Using the definition~\eqref{eq:FreeFieldOperator} (with $\hat\psi_{\rm free}\equiv\sum_h\hat\psi_{h,\mathrm{free}}$), decompositions~\eqref{eq:decompose} and spatial translation invariance, this can be written as
\begin{equation}
    \overbar S_{hh'\evec{k}}^>(k_0,t)
    = \delta_{hh'} \hspace{-.3em} \int_{-\infty}^{\infty}{\rm d}r_0\,\e^{\im k_0r_0}
    M^>_{h\evec{k}}\bigl( t + \sfrac{r_0}{2}, t - \sfrac{r_0}{2}\bigr) \otimes
    P^{\scriptscriptstyle (2)}_{h\evec{k}} \text{,}
\label{eq:plotted}
\end{equation}
where $P^{\scriptscriptstyle (2)}_{h\evec{k}} = \xi_{h\evec{k}}\xi_{h\evec{k}}^{\dag} = \frac{1}{2}(\idmat + h\evec{\sigma} \cdot \hat{\evec{k}})$ and only the chiral component matrix $M^>_{h\evec{k}}$ depends on the mode functions:
%
\begin{equation}
M^>_{h\evec{k}}\bigl(t+\sfrac{r_0}{2},t-\sfrac{r_0}{2}\bigr) \equiv
\frac{1}{2\omega_{-}}\begin{bmatrix}
\eta_{h\evec{k}}\bigl(t+\frac{r_0}{2}\bigr)\eta_{h\evec{k}}^*\bigl(t-\frac{r_0}{2}\bigr) & \hspace{0.3em}\eta_{h\evec{k}}\bigl(t+\frac{r_0}{2}\bigr)\zeta_{h\evec{k}}^*\bigl(t-\frac{r_0}{2}\bigr) \\
\zeta_{h\evec{k}}\bigl(t+\frac{r_0}{2}\bigr)\eta_{h\evec{k}}^*\bigl(t-\frac{r_0}{2}\bigr) & \hspace{0.3em}\zeta_{h\evec{k}}\bigl(t+\frac{r_0}{2}\bigr)\zeta_{h\evec{k}}^*\bigl(t-\frac{r_0}{2}\bigr)
\end{bmatrix} \text{.}
\label{eq:plotted_M}
\end{equation}
When the component mode functions are solved, it is straightforward to construct the Wightman function using fast Fourier transform methods.

%
\subsection{Including damping}
%

In the absence of dissipative processes, the free particle solutions~\eqref{eq:plotted} are correlated over arbitrarily large time intervals, because the Wigner transform correlates mode functions over all relative times $\pm\frac{r_0}{2}$ at each value of $t$. This is of course a physical result. However, our typical applications concern interacting systems, where such correlations are naturally suppressed by decohering interactions.

Taking interactions completely into account would require solving the full Kadanoff--Baym equations, which is beyond the scope of this paper. However, one can account for their most important effect for the phase space structure in a rather simple manner. We observe that the information encoded in the relative coordinate must be damped by the rate of interactions that measure the state of the system (in this case whether the system is a particle or an antiparticle). If we denote this rate by $\Gamma_{h\evec{k}}$ for each mode with momentum $\evec{k}$ and helicity $h$, then the appropriately damped correlation function should be
\begin{align}
    \overbar S^{>}_{h\evec{k},\Gamma}(k_0,t)
    &\equiv \int {\rm d}^4r\,\e^{\im k_0r_0-\im\evec{k}\cdot\evec{r} - \Gamma_{h\evec{k}} |r_0|} 
    \bigl\langle\Omega\bigl|
        \hat \psi_{h,\rm free} \bigl(x+\sfrac{r}{2}\bigr) \hat {\overbar{\psi}}_{h,\rm free} \bigl(x-\sfrac{r}{2}\bigr)
    \bigr|\Omega\bigr\rangle \gamma^0 \notag \\
    &= \int_{-\infty}^{\infty}{\rm d}r_0\,\e^{\im k_0r_0 - \Gamma_{h\evec{k}} |r_0|}
    M^>_{h\evec{k}}\bigl(t+\sfrac{r_0}{2},t-\sfrac{r_0}{2}\bigr)\otimes P^{\scriptscriptstyle (2)}_{h\evec{k}} \notag \\
    &\equiv W^{>}_{h\evec{k},\Gamma}(k_0,t) \otimes P^{\scriptscriptstyle (2)}_{h\evec{k}}\text{.}
    \label{eq:DampedWightmanFunction}
\end{align}
The only difference to the exact free case~\eqref{eq:plotted} is the introduction of the exponential damping factor $\e^{-\Gamma_{h\evec{k}} |r_0|}$, where the damping rate $\Gamma_{h\evec{k}}$ is the imaginary part of the pole of the full propagator. The exponential accounts for the most relevant effect of interactions here. Taking the self-energy fully into account would also modify the matrix $M^>_{h\evec{k}}$, which we here approximate with the free result. Equation~\eqref{eq:DampedWightmanFunction} is thus reasonable in the usual weak coupling limit, where particles are assumed to propagate freely between relatively infrequent collisions.\footnote{In fact we are accounting also for the soft interactions with the background fields that lead to the time-varying mass term. It would be straightforward to extend this to other dispersive processes by the use of quasiparticle eigenstates.} When collisions occur they affect ``measurements'' of the quantum state, which over time leads to a loss of coherence.

The appearance of the exponential damping factor in equation~\eqref{eq:DampedWightmanFunction} can also be motivated by studying the case of thermal equilibrium, where the full correlation function in Wigner representation is given by $\overbar S^>_{h\evec{k}}(k_0,t) = 2\overbar{\mathcal{A}}_{h\evec{k}}(k_0,t)(1 - f_\mathrm{eq}(k_0))$. (Remember that $\overbar{S}^>_{h\evec{k}} + \overbar{S}^<_{h\evec{k}} = 2\overbar{\mathcal{A}}_{h\evec{k}}$). The damping factor in this case arises from the absorptive self-energy corrections to the single particle poles of the pole propagators $S_{h\evec{k}}^{\rm r,a}$. When neglecting gradient corrections one can show that in the small coupling limit
\begin{gather}
    \overbar S_{h\evec{k}}^>(k_0,t)
    \simeq \int\mathrm{d}r_0\,\e^{\im k_0r_0 - \Gamma_{h\evec{k}}(t)|r_0|}
        \overbar S_{0,h\evec{k}}^>\bigl(t+\sfrac{r_0}{2},t-\sfrac{r_0}{2}\bigr) \text{,}
    \label{eq:DampedWightmanFunctionExample} \\
\shortintertext{where}
    \overbar S_{0,h\evec{k}}^>\bigl(t + \sfrac{r_0}{2}, t - \sfrac{r_0}{2}\bigr)
    = \sum_{\pm}\e^{\mp\im \omega_\evec{k}(t)r_0}\Bigl[1 - f_\mathrm{eq}\bigl(\pm\omega_\evec{k}(t)\bigr)\Bigr]
        P^{\scriptscriptstyle (4)}_{h\evec{k}} P^\pm_{\evec{k}}(t)
    \label{eq:DampedWightmanFunctionExample_b}
\end{gather}
is the two-time representation of the free thermal correlation function (derived using the usual plane wave mode functions). We have only kept the absorptive corrections to the single particle poles of $S^{\mathrm{r},\mathrm{a}}_{h\evec{k}}(k_0,t)$, which are then located at $k_0 = \omega_\evec{k}(t) \mp \im\Gamma_{h\evec{k}}(t)$. The damping factor in equation~\eqref{eq:DampedWightmanFunction} relates the free correlation function to the full one in exactly the same way as in equation~\eqref{eq:DampedWightmanFunctionExample}, generalising the latter into the case of a non-thermal system with coherence structures.

%
\subsection{Explicit solutions for mode functions}
\label{sec:phase}
%

\begin{figure}[t!]
  \centering
  \begin{subfigure}[b]{0.5\linewidth}
    \hspace{.5em}\centering\includegraphics[width=225pt]{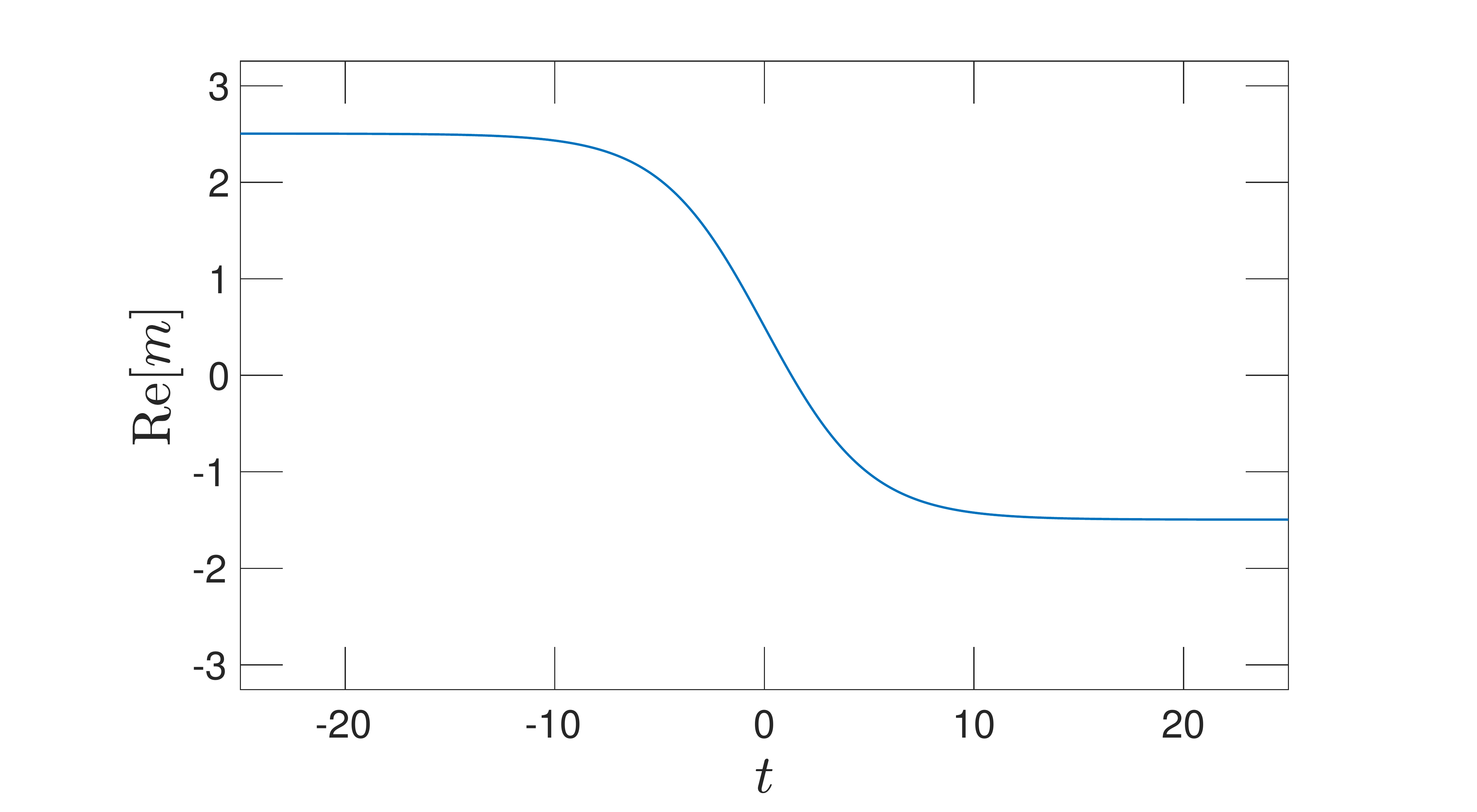}
    \caption{\label{fig:tanh1}}
  \end{subfigure}%
  \begin{subfigure}[b]{0.5\linewidth} 
    \hspace{-1.1em}\centering\includegraphics[width=225pt]{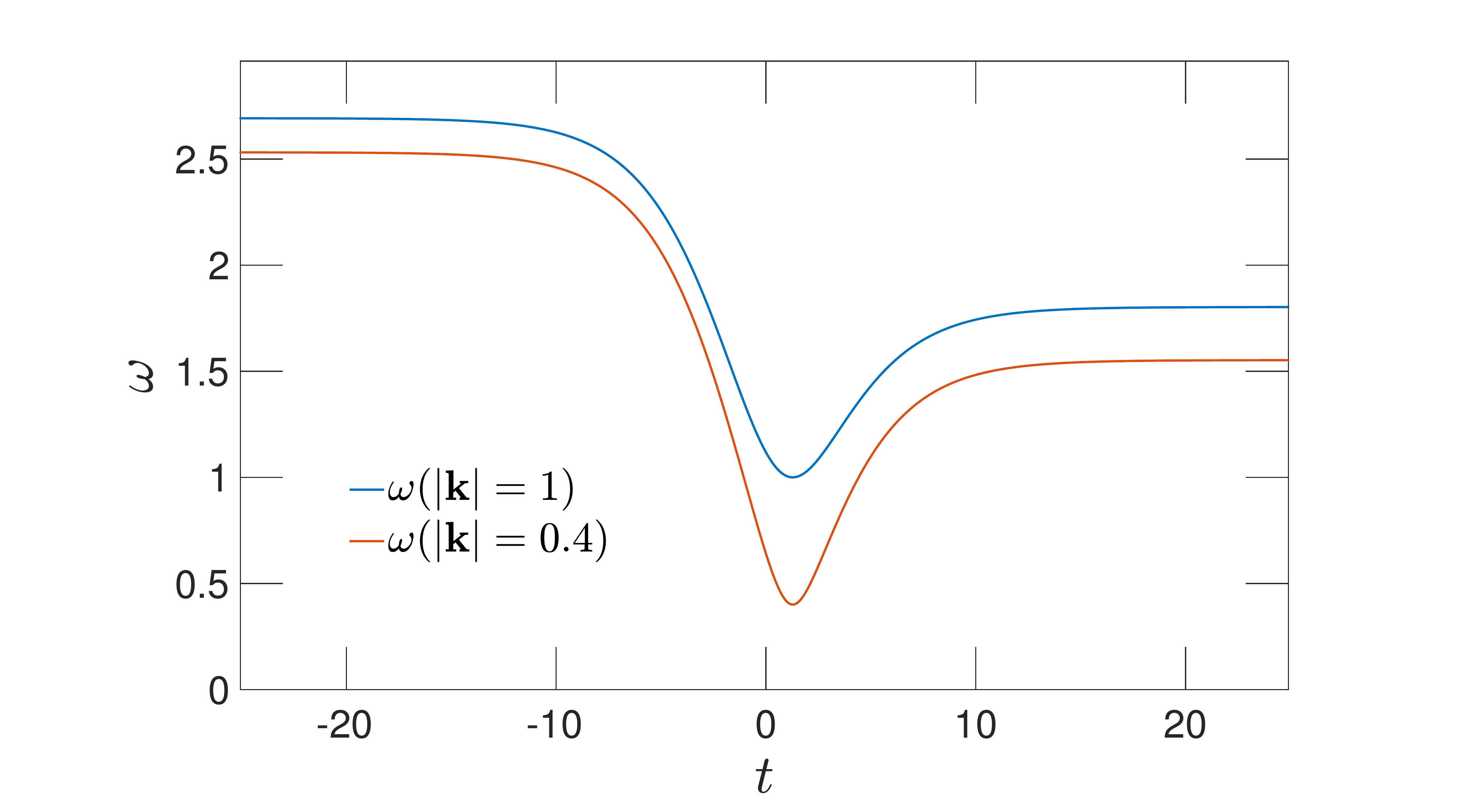}
    \caption{\label{fig:tanh2}}
  \end{subfigure}
\caption{Left panel (a): the real part of the mass profile $m(t)$ of equation~\eqref{kink}, with parameters $m_1 = 0.5 + 0.005 \im$, $m_{2} = 2$ and $\tau_{\rm w} = 5$, in arbitrary units. Right panel (b): the positive energy eigenvalue $\omega_\evec{k}(t)=\sqrt{\evec{k}^2+|m(t)|^2}$ with the same mass function as in the left panel and with both $|\evec{k}|= 1$ and $0.4$.}
\label{fig:tanh}
\end{figure}

We shall now study the correlation function~\eqref{eq:DampedWightmanFunction} 
explicitly in a simple toy model.
For quantitative results we must define the mass function $m(t)$. We assume that it approaches asymptotically constant values $m_\mp$ at early and late times, respectively, and that it changes between the asymptotic values over a characteristic time interval $\tau_{w}$ around time $t=0$. This is the situation \eg in a phase transition interpolating between the broken and unbroken phases. At early and late times such solutions approach asymptotically plane waves (with spinor normalisation $U_{h\evec{k}}^\dagger U_{h\evec{k}} = V_{h\evec{k}}^\dagger V_{h\evec{k}} = 2\omega_{-}$):
\begin{subequations}
\label{eq:ConstantMassSolutinos}
\begin{align}
\label{eq:ConstantMassSolutinosa}
U^\infty_{h\evec{k}} &= \begin{bmatrix*}[l] 
    \sqrt{\omega_{-}-h|{\evec{k}}|} \\
    \mathrlap{\sqrt{\omega_{-}+h|{\evec{k}}|}\e^{-\im\theta}}
    \phantom{-\sqrt{\omega_{-}-h|{\evec{k}}|}\e^{\im\theta}}
\end{bmatrix*} \otimes \xi_{h\evec{k}} \e^{-\im \omega_{-} t} \text{,} \\
\label{eq:ConstantMassSolutinosb}
V^\infty_{h\evec{k}} &= \begin{bmatrix*}[l]
    \phantom{-}\sqrt{\omega_{-}+h|{\evec{k}}|} \\
    -\sqrt{\omega_{-}-h|{\evec{k}}|}\e^{\im\theta}
\end{bmatrix*} \otimes \xi_{h\evec{k}} \e^{\im \omega_{-} t} \text{,}
\end{align}
\end{subequations}
where $\theta$ is the phase of the constant mass in the asymptotic limit: $m \rightarrow |m_\pm|\e^{\im\theta_\pm}$. To be specific, we use the following mass profile for which the mode functions can be solved analytically~\cite{Prokopec:2013ax}:
\begin{equation}
\label{kink}
    m(t) = m_1 + m_2\tanh\biggl(-\frac{t}{\tau_{\rm w}}\biggr) \text{,}
\end{equation}
where $m_1 = m_{1\mathrm{R}} + \im m_{1\mathrm{I}}$ and $m_2 = m_{2\mathrm{R}} + \im m_{2\mathrm{I}}$ are constant complex coefficients and $\tau_{\rm w}$ is a parameter describing the width of the transition in time. At early times ($t \rightarrow -\infty$) we then have $m \rightarrow m_{-} = m_1+m_2$ and at late times ($t \rightarrow \infty$) $m\rightarrow m_{+} = m_1-m_2$. For solving the mode functions, the imaginary part of $m_2$ is removed by a global rotation of the spinors (see ref.~\cite{Prokopec:2013ax} for details), which of course does not change the dynamics of the system. The remaining imaginary part is simply denoted by $m_{\mathrm{I}}$. Figure~\ref{fig:tanh} illustrates the shape of the mass function and the corresponding energy for representative parameters.

%
%
%
%

Equations~\eqref{eq:ChiralComponentEom} with the mass profile~\eqref{kink} were solved in ref.~\cite{Prokopec:2013ax} and here we just quote the results relevant for our purposes. Defining a new basis for the mode functions,
\begin{subequations}
\label{pmbasis}
\begin{align}
    \phi_{h\evec{k}}^{\pm}(t) &\equiv \frac{1}{\sqrt{2}}\bigl[
        \eta_{h\evec{k}}(t)\pm\zeta_{h\evec{k}}(t)
    \bigr]\text{,} \\
    \overbar{\phi}_{h\evec{k}}^{\pm}(t) &\equiv  \frac{1}{\sqrt{2}}\bigl[
        \overbar{\eta}_{h\evec{k}}(t)\pm\overbar{\zeta}_{h\evec{k}}(t)
    \bigr]\text{,}
\end{align}
\end{subequations}
one can show that the solutions can be written in terms of Gauss' hypergeometric functions:
\begin{subequations}
\label{phisol}
\begin{align}
\label{phisola}
\phi_{h\evec{k}}^{\pm(1)} &= C_{h\evec{k}}^{\pm(1)}\,z^{\alpha}(1-z)^{\beta}\,
    \prescript{}{2}{F}^{}_{1}(a_{\pm}, b_{\pm}, c; z)\text{,} \\
\label{phisolb}
\phi_{h\evec{k}}^{\pm(2)} &= C_{h\evec{k}}^{\pm(2)} \,z^{-\alpha}(1-z)^{\beta}\,
    \prescript{}{2}{F}^{}_{1}(1+a_{\pm}-c, 1+b_{\pm}-c, 2-c; z)\text{,}
\end{align}
\end{subequations}
where $C_{h\evec{k}}^{\pm(1,2)}$ are constants and
\begin{equation}
\begin{split}
z &= \frac{1}{2}\left[1-\tanh\left(-\frac{t}{\tau_{\rm w}}\right)\right]\text{,} \qquad \alpha = -\frac{\im}{2}\tau_{\rm w}\omega_{-}\text{,} \qquad \beta = -\frac{\im}{2}\tau_{\rm w}\omega_{+}\text{,} \\
\omega_{\mp} &= \sqrt{\evec{k}^2+m_{\mathrm{I}}^2+(m_{1\mathrm{R}}\pm m_{2\mathrm{R}})^2}\text{,} \\
a_{\pm} &\equiv 1+\alpha+\beta\mp\im\tau_{\rm w}m_{2\mathrm{R}}\text{,} \qquad b_ {\pm} \equiv \alpha+\beta\pm\im\tau_{\rm w}m_{2\mathrm{R}}\text{,} \qquad c \equiv 1+2\alpha\text{.}
\end{split}
\end{equation}
Superscripts $(1)$ and $(2)$ label the two linearly independent solutions. The solutions for $\overbar{\phi}_{h\evec{k}}^{\pm}$ can be obtained by changing the sign of helicity in equations~\eqref{phisol}, $h \rightarrow -h$. (Helicity enters the solution through the boundary conditions as will be seen below.)

Using the properties of the hypergeometric functions it is easy to check that at early times
\begin{equation}
    \phi_{h\evec{k}}^{\pm(1)}\,\xrightarrow{t\rightarrow -\infty}\,
        C_{h\evec{k}}^{\pm(1)}\e^{-\im t\omega_{-}}\text{,} \hspace{3em}
    \phi_{h\evec{k}}^{\pm(2)}\,\xrightarrow{t\rightarrow -\infty}\,
        C_{h\evec{k}}^{\pm(2)}\e^{\im t\omega_{-}}\text{.}
\end{equation}
At late times these solutions split into mixtures of positive and negative frequency states:
\begin{equation}
    \phi_{h\evec{k}}^{\pm(1)} \, \xrightarrow{t\rightarrow \infty}\,
    C_{h\evec{k}}^{\pm(1)}\,\frac{\Gamma(c)\Gamma(c-a_{\pm}-b_{\pm})}{\Gamma(c-a_{\pm})\Gamma(c-b_{\pm})}\e^{\im t\omega_{+}} + C_{h\evec{k}}^{\pm(1)}\,\frac{\Gamma(c)\Gamma(a_{\pm}+b_{\pm}-c)}{\Gamma(a_{\pm})\Gamma(b_{\pm})}\e^{-\im t\omega_{+}}\text{,}
\end{equation}
which manifests the fact that a varying mass mixes particle and antiparticle states. Indeed, in systems without time-translation invariance the division to particles and antiparticles is not unique. Locally a clear identification can be made however, and with the asymptotic limits given above we can construct different initial and final states we wish to study.

\begin{figure}[t!]
\vspace{-1em}
\hspace{-2.5em}
\includegraphics[scale=0.36]{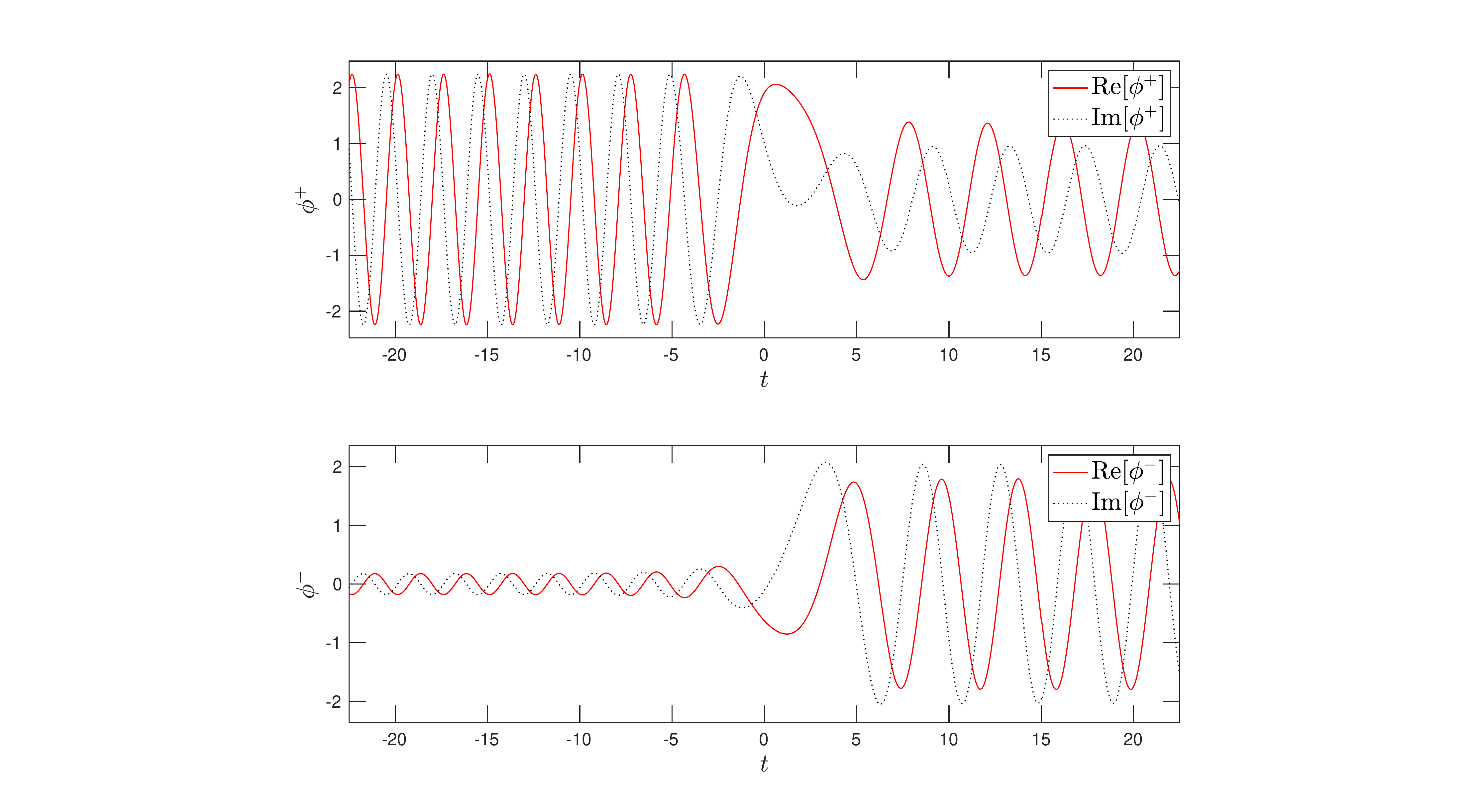}
\caption{Shown are the real and imaginary parts of the exact free mode functions $\phi_{h\evec{k}}^{\pm(1)}$, defined in equation~\eqref{phisola}, across the transition defined by the mass profile~\eqref{kink}. We used the initial conditions~\eqref{eq:boundarycond} and the same parameters as in figure~\ref{fig:tanh1} with $\big|\evec{k}\big| = 0.4$ and $h = 1$.}
\label{fig:hypgeom}
\end{figure}

Let us now specify our initial state as a positive frequency particle, \ie the solution~\eqref{phisola}, corresponding to the constant mass one-particle state~\eqref{eq:ConstantMassSolutinosa} at $t \rightarrow -\infty$. This determines the constants
\begin{equation}
    C_{h\evec{k}}^{\pm(1)} = \frac{1}{\sqrt{2}}\biggl(
        \sqrt{\omega_{-} -h|\evec{k}|} \pm \sqrt{\omega_{-} +h|\evec{k}|}\e^{-\im\theta_-}
    \biggr)\text{,}
\label{eq:boundarycond}
\end{equation}
where $\theta_- = \operatorname{Arg}(m_{1\mathrm{R}} + m_{2\mathrm{R}} + \im m_{\mathrm{I}})$. Figure~\ref{fig:hypgeom} shows these solutions for a representative set of parameters. It is evident that the solutions asymptote to plane waves very quickly on each side of the transition region.

%
%
%
%

%
\section{Phase space of the exact Wightman function}
\label{sec:phasespace}
%

\begin{figure}[t!]
\begin{center}
  \includegraphics[width=0.49\textwidth]{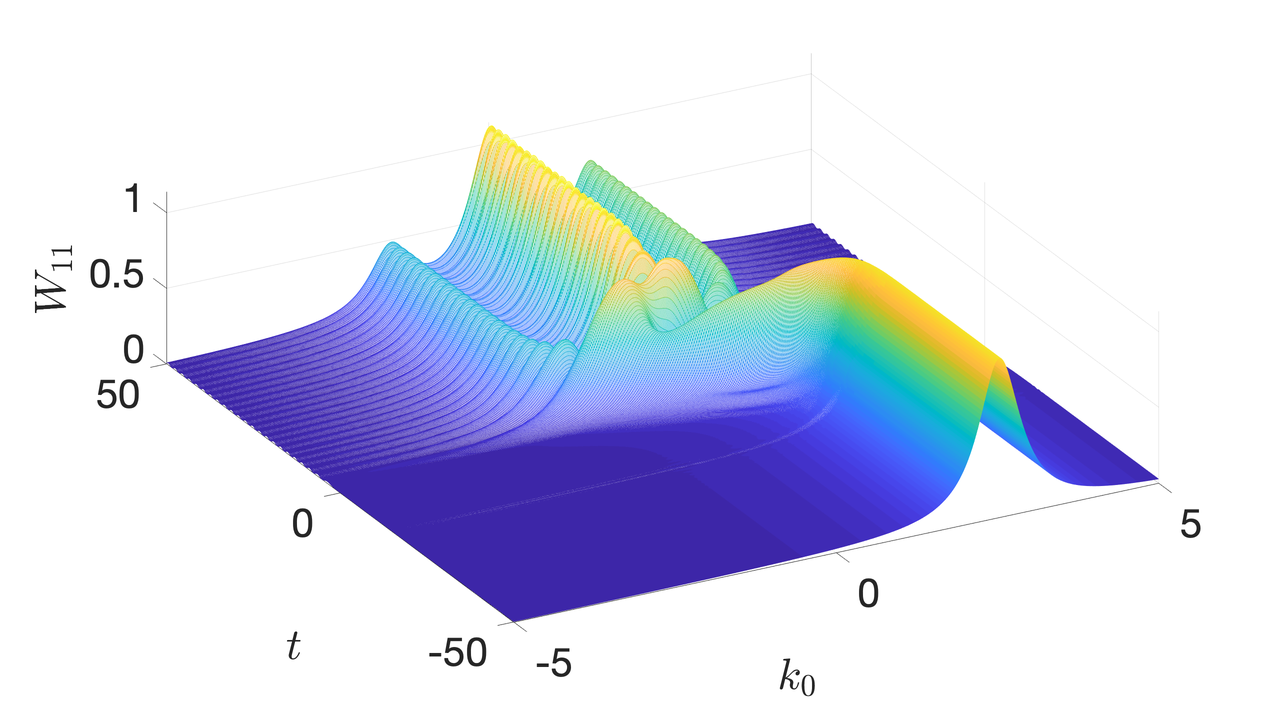} 
  \includegraphics[width=0.49\textwidth]{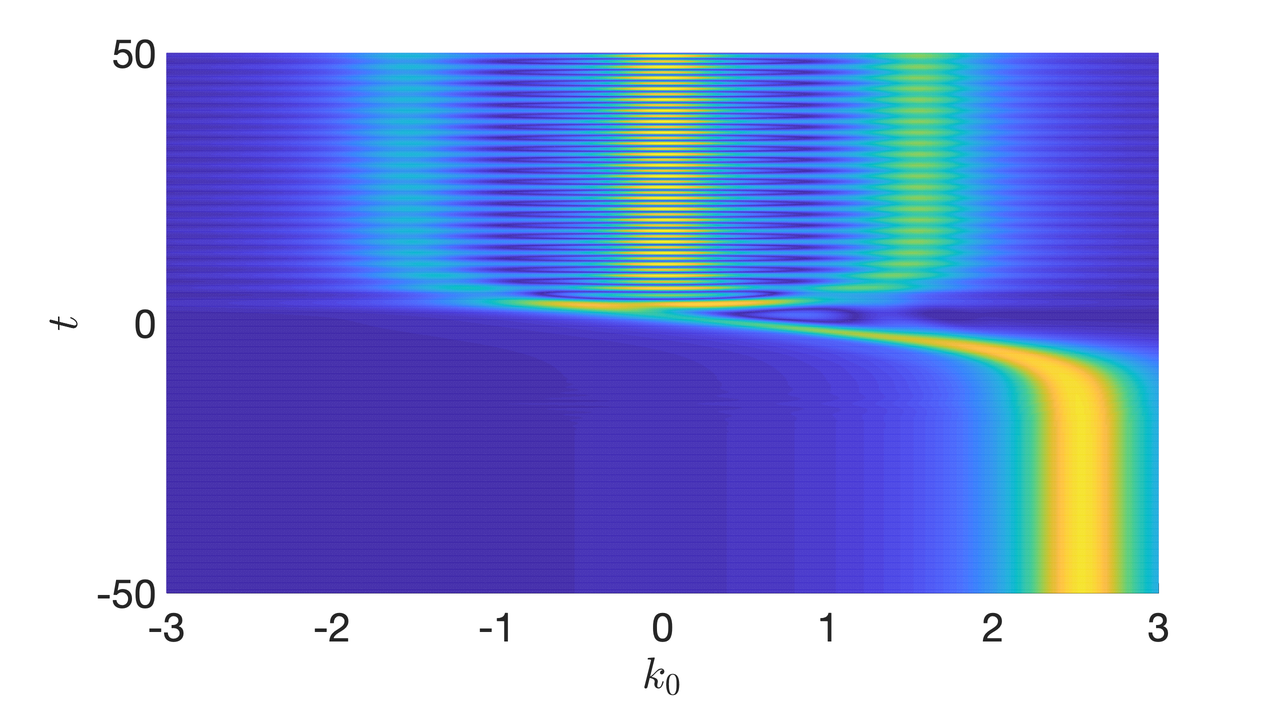} 
%
\caption{Shown is the absolute value of the $(1,1)$-component of the exact free Wightman function $W^>_{h\evec{k},\Gamma}$ defined in equation~\eqref{eq:DampedWightmanFunction}, for parameters $h=1$, $|\evec{k}|=0.4$, $m_{1\mathrm{R}}=0.5$, $m_{2\mathrm{R}}=2$, $m_{\mathrm{I}}=-0.005$, $\tau_{\rm w}=5$ and $\Gamma=0.4$. Note that time flows from bottom to top in the right panel.}
\label{fig:iS_largeGamma}
\end{center}
\end{figure}

\begin{figure}[t!]
\begin{center}
    \includegraphics[width=0.49\textwidth]{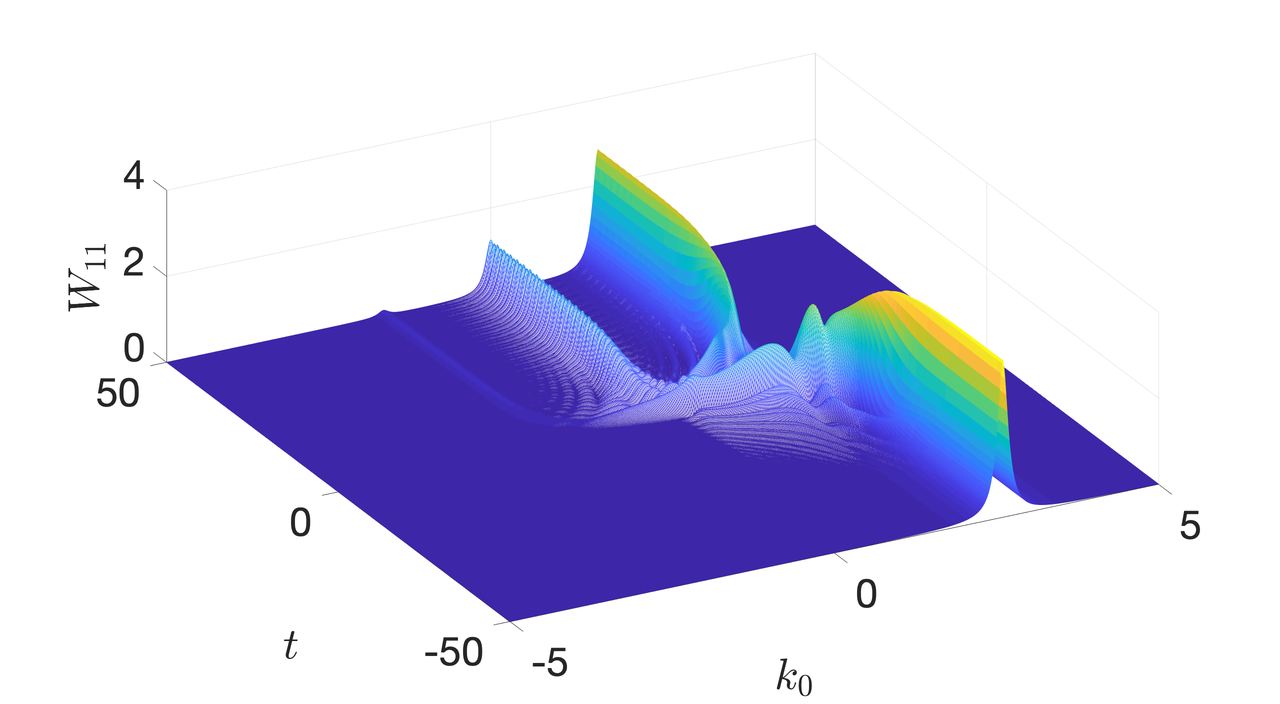} 
    \includegraphics[width=0.49\textwidth]{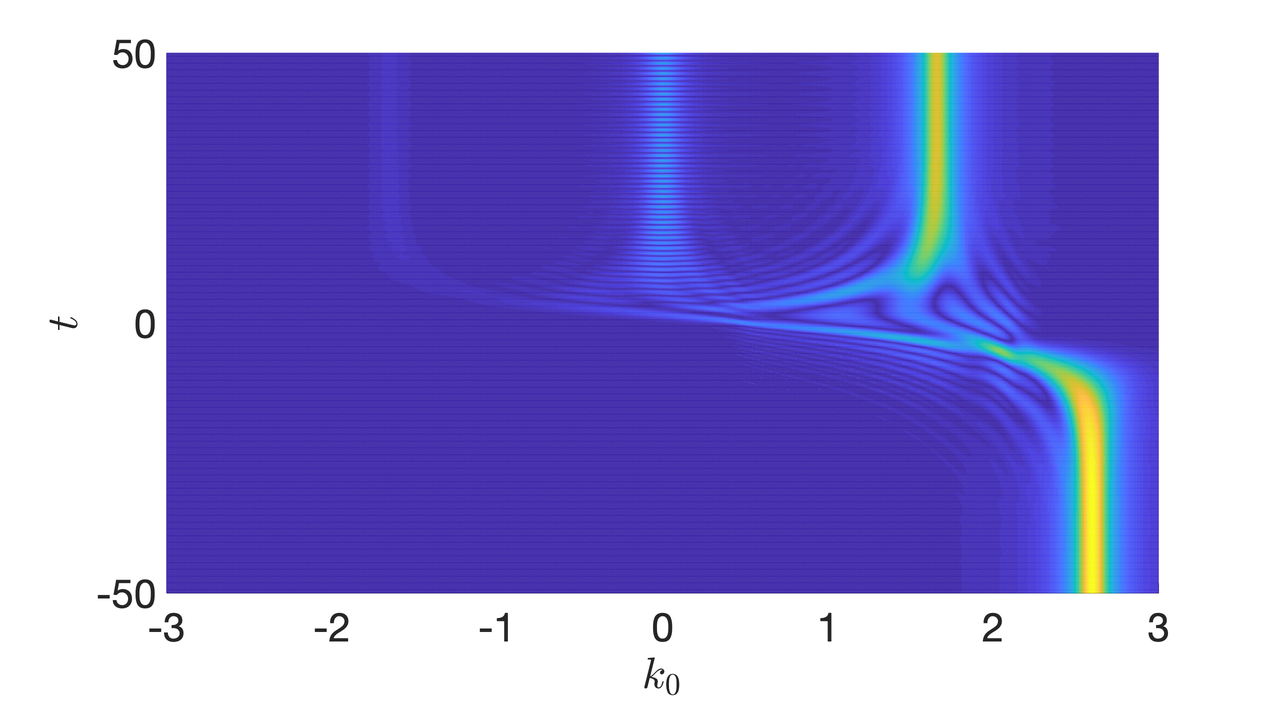} 
%
    \caption{The same as in figure~\ref{fig:iS_largeGamma}, but for parameters $h=1$, $|\evec{k}|=0.7$, $m_{1\mathrm{R}}=0.5$, $m_{2\mathrm{R}}=2$, $m_{\mathrm{I}}=-0.005$, $\tau_{\rm w}=5$ and $\Gamma=0.1$.}
    \label{fig:iS_largeGamma_largetau}
\end{center}
\end{figure}

Having solved the mode functions, we can now calculate the Wightman functions $S_{h\evec{k}}^{s}$ and $S_{h\evec{k},\Gamma}^{s}$. It suffices to concentrate on one type of them, say $S^>$, since both functions exhibit the same phase space structures. We evaluate the Wightman functions by inserting the mode functions solved from equations~\cref{pmbasis,phisol} with the boundary conditions~\eqref{eq:boundarycond} into the matrix $M_{h\evec{k}}^>$~\eqref{eq:plotted_M} and performing the integral over the relative coordinate in equation~\eqref{eq:DampedWightmanFunction} numerically for each $\evec{k}$-mode. Results of these computations for varying parameter sets are shown in figures~\cref{fig:iS_largeGamma,fig:iS_largeGamma_largetau,fig:iS_smallGamma}.

Figure~\ref{fig:iS_largeGamma} shows the absolute value of the $(1,1)$-component of the function $W^>_{h\evec{k},\Gamma}(k_0,t)$, defined in equation~\eqref{eq:DampedWightmanFunction}, for a system initially prepared to a pure positive frequency state. (Other three chiral components are qualitatively similar.)  The surface plot in the left panel displays a clearly peaked structure, where the initial particle peak branches at the transition region to three separate peaks corresponding to particle and antiparticle solutions at $k_0 = \pm \omega_\evec{k}(t)$ and a coherence peak at $k_0 = 0$. This reproduces the cQPA-shell structure predicted in the previous section. Note that the coherence shell solution is rapidly oscillating in time as predicted by the cQPA equation~\eqref{eq:free_equations3}. The feature is slightly obscured by the absolute value, but it shows up in the ``digitised'' structure of the coherence solution in the projected plot on the right panel. Due to a rather large interaction rate $\Gamma$ the shell structures are wide enough in frequency to overlap a little, which can after the transition be seen as a leakage of the coherence shell oscillations into the mass shells. At early times the sole positive frequency shell contains no oscillations. Physically, what we are seeing, is particle production by a temporally changing mass parameter and the fundamental relation of the phenomenon to the quantum coherence between positive and negative frequency states.

%
%
%
%

In figure~\ref{fig:iS_largeGamma} we assumed a quite large damping factor and correspondingly the shell structures were rather broad in frequency. In figure~\ref{fig:iS_largeGamma_largetau} we show for comparison a solution with a smaller wavelength and a much smaller damping coefficient. As expected, the shell structure gets more sharply peaked because of the smaller width.\footnote{In fact it is easy to show in an even simpler toy model, where the mass-function is replaced by a step-function, that the peaks become Breit--Wigner-functions in frequency~\cite{Koskivaara:2015tcc}. The spectral cQPA-solution can then be seen explicitly as the Breit--Wigner forms approach delta functions in the limit $\Gamma \rightarrow 0$.} At the same time the antiparticle shell after the transition becomes much less pronounced, reflecting the fact that a larger initial energy is less affected by the mass change. (The same qualitative behaviour would of course be obtained by increasing the width of the wall, leading to less efficient particle production.) Indeed, for a very large $|\evec{k}|$ the whole novel shell structure vanishes, making way for a single shell following a classical energy path such as the ones shown in figure~\ref{fig:tanh2}.

%
%
%
%

Right at the transition region one can distinguish additional fine-structures, which are not related to the cQPA solution of equation~\eqref{eq:cQPA-correlator-free-particle}. This is partly because our derivation of cQPA assumed lowest order expansion in gradients. It would be interesting (and possible) to generalise cQPA to a singular higher order expansion in gradients and check if the emerging discrete sequence of shells could reproduce the structures seen here. However, these structures may also reflect the onset of the new non-local correlations that we shall turn to next.\footnote{Let us clarify our use of the notion of (non-)locality in this paper: first, by non-local coherence we mean coherence over the relative coordinate in the two-point correlation function. Then, by local limit, we mean the limit where the two time-arguments in the correlation function are the same. The local correlation function still supports the particle-antiparticle coherence, which is non-local in the sense that creating it requires coherent evolution over a finite interval in the average time uninterrupted by collisions, which differentiate particles from antiparticles.}

%
\subsection{Non-local coherence in time}
%

\begin{figure}[t!]
\begin{center}
   \includegraphics[width=0.49\textwidth]{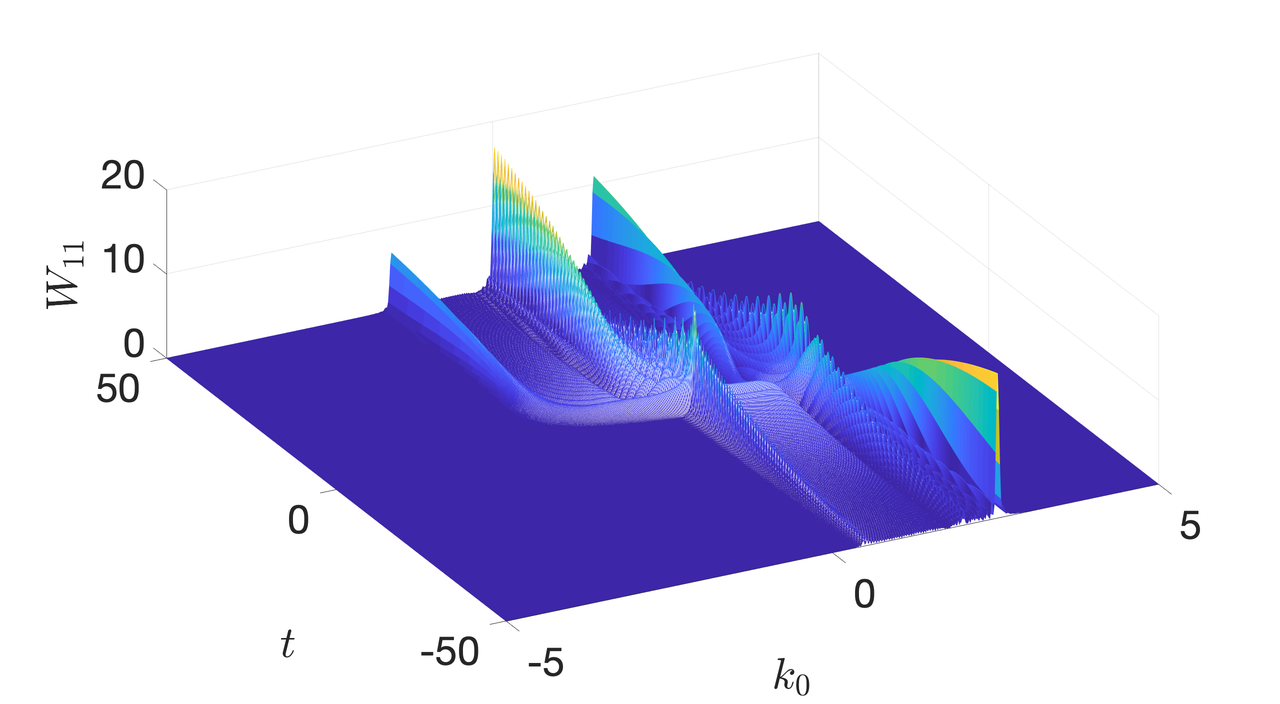}
   \includegraphics[width=0.49\textwidth]{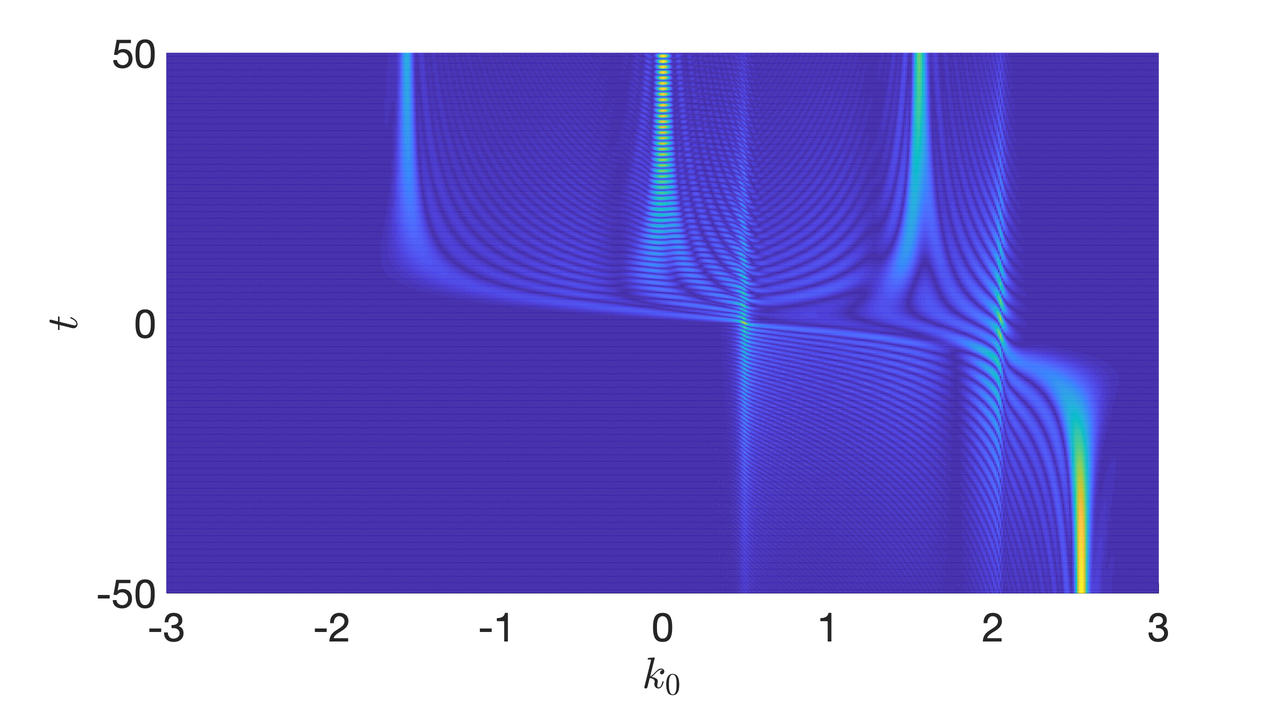}
\caption{The same as in figure~\ref{fig:iS_largeGamma}, but for parameters $h=1$, $\bigl|\evec{k}\bigr|=0.4$, $m_{1\mathrm{R}}=0.5$, $m_{2\mathrm{R}}=2$, $m_{\mathrm{I}}=-0.005$, $\tau_{\rm w}=5$ and $\Gamma=0.02$.}
\label{fig:iS_smallGamma}
\end{center}
\end{figure}

In figure~\ref{fig:iS_smallGamma} we again plot $|W^>_{h\evec{k},\Gamma,11}|$ with the same parameters as in figure~\ref{fig:iS_largeGamma}, but with a much smaller decay term. The shells become even more peaked as expected, but in addition a much richer phase space structure emerges, extending well outside the transition region. From the projection plot one recognises that two new spectral shells have entered the play, together with a rich network of secondary fine-structures around the transition region. From the surface plot it is evident, compared to the earlier cases, that the cQPA-shells are suppressed near the transition region, while the new shells grow in amplitude there. Far away from the transition region the situation is reversed and the new shells (which are also oscillating) fade away, making room for the usual cQPA-shells that allow for a clear particle and antiparticle identification.

%
%
%
%

The new shells correspond to non-local correlations between the early- and late-time solutions across the wall; in the particle interpretation the system appears to become aware of the change in its energy levels already before the transition occurs. This is completely expected behaviour for a quantum system and, again, these shells can also be seen analytically in the simple step-function model~\cite{Koskivaara:2015tcc}. One can show, and also observe in the projection plot, that the new shells coincide with the average frequencies
\begin{equation}
    k_0 = \frac{1}{2}\biggl(
        \sqrt{\evec{k}^2 + |m_{-}|^2} \pm \sqrt{\evec{k}^2 + |m_{+}|^2}
    \biggr) \text{,}
\end{equation}
which reveals that they correspond to particle-particle and particle-antiparticle correlations across the wall. The reason why these solutions are suppressed at large time differences is the damping; the information about the transition can be propagated only up to a distance $\Delta t \sim 1/\Gamma$ in the relative coordinate. Beyond this time interval only local correlations can survive. Decreasing $\Gamma$ further makes the non-local coherence structures ever more prominent and if one removes damping entirely, the system becomes completely overwhelmed by them. In this limit the system is intrinsically quantum; local particle-like solutions are irrelevant and the system is globally sensitive to the initial conditions and the size of the time-domain.

%
\subsection{Physical and practical significance of the phase space structures}
%

We have seen that a quantum system with negligible damping is strongly correlated over large time intervals. However, in interacting systems damping suppresses non-local correlations, eventually reducing correlation functions to the local limit. This decoherence enables the quasiparticle picture and eventually the Boltzmann limit in slowly varying backgrounds. In the language of a direct space Kadanoff--Baym approach, damping removes contributions from memory integrals over long relative time differences. Note however, that damping does \emph{not} destroy the coherence shell at $k_0=0$; spectral cQPA shells get finite widths, but the coherence between particles and antiparticles survives. Of course, equations~\cref{PoleEqMix2,StatEqMix2} contain also other (hard) collisions terms, which we have omitted so far. If these collisions depend on the particle-antiparticle nature of the state, they constitute measurements which destroy this coherence. A complete treatment of particle production in phase transitions, for example, should account for this effect as well, as was indeed done for example in refs.~\cite{Herranen:2008hu,Herranen:2008di} in the cQPA context.

From a practical point of view our solutions show that in the weakly interacting limit $\tau_{\rm w}\Gamma\ll1$, a complete phase space solution of the interacting problem would require very fine resolution in frequency space in order to account for all the fine-structures in the transition region. In this region, because of the large number of transient shell structures, the quasiparticle picture appears impractical.\footnote{The situation is not as bad as one might think even in the limit $\tau_{\rm w} \Gamma \ll 1$. Let us consider the problem from the point of view of the cQPA method, which includes local coherence shells but ignores the non-local structures. Because the quasiparticle picture is appropriate far from the transition region and one expects only few interactions within the transition area, the evolution of the quasiparticle distributions may be only weakly sensitive to the new transient structures (the evolution of the quasiparticle functions is affected by the new shells only through the collision integrals). If the physics one is interested in is sensitive only to the late time correlations, it should be rather well described by the cQPA.} On the other hand, even for a moderately strongly interacting system $\tau_{\rm w} \Gamma \gtrsim 0.5$, the phase space structure is smoothed out and the \emph{coherent} quasiparticle picture of refs.~\cite{Herranen:2008hi,Fidler:2011yq,Herranen:2008hu,Herranen:2008di,Herranen:2009zi,Herranen:2009xi,Herranen:2010mh} should provide a good description of the system.

%
\section{Currents and connection to the semiclassical limit}
\label{sec:currents}
%

In the previous sections we showed that the phase space of a system with a varying mass profile has non-trivial phase space structures, whose intricacy depends on the size of the mode momentum $\evec{k}$ and the damping strength $\Gamma$. We also argued that the quasiparticle picture may provide a reasonable description of the system (even for very small $\tau_{\rm w}\Gamma$). We now change slightly our perspective, and ask how our results compare with the semiclassical treatment, which should be applicable when $\tau_{\rm w}|\evec{k}|\gg 1$. Semiclassical methods have been widely used to describe CP-violating dynamics in electroweak baryogenesis models~\cite{Cline:2000nw,Cline:2001rk,Kainulainen:2001cn,Kainulainen:2002th,Prokopec:2003pj,Prokopec:2004ic,Fromme:2006cm,Fromme:2006wx,Cline:2012hg,Cline:2013gha,Cline:2017qpe}. While we are dealing with a purely time-dependent system here, the results should be qualitatively representative. 

To be specific, we shall compare different methods for computing the expectation values of fermionic currents. A generic current corresponding to a Dirac operator $\mathcal{O}$ can be computed as 
\begin{equation}
\label{eq:current}
    j^{\mathcal{O}}_{h\evec{k}}(t)
    \equiv \int \frac{{\rm d}^3\evec{r}}{(2\uppi)^3}\,\e^{\im\evec{k}\cdot\evec{r}}\bigl\langle
        \hat{\overbar{\psi}}_{h}\bigl(t,\evec{x}+\evec{r}\bigr) \mathcal{O}
        \hat\psi_{h}\bigl(t,\evec{x}\bigr)
    \bigr\rangle
    = \int\frac{\mathrm{d}k_0}{2\uppi}\Tr\bigl[
        \mathcal{O}\,\im S_{{h\evec{k}}}^<(k_0,t)
    \bigr] \text{.}
\end{equation}
In particular, we will be interested in the axial charge density
\begin{equation}
    j_{5,h\evec{k}}(t) \equiv
    \int \frac{{\rm d}^3\evec{r}}{(2\uppi)^3}\,\e^{\im\evec{k}\cdot\evec{r}}\bigl\langle
        \hat{\overbar{\psi}}_{h}\bigl(t,\evec{x}+\evec{r}\bigr) \gamma^0\gamma^5
        \hat\psi_{h}\bigl(t,\evec{x}\bigr)
    \bigr\rangle \text{,}
    \label{eq:axialDensity}
\end{equation}
which is related to particle asymmetries.

With the exact solutions~\eqref{phisol} at hand it is a simple numerical task to compute $j_{5,h\evec{k}}$ for the kink profile using equation~\eqref{eq:current}. Furthermore, in cQPA it can be calculated in terms of the shell functions $f_{h\evec{k}}^{{(m,c)\pm}}$ as follows:
\begin{equation}
\label{eq:cqpatosc}
    j_{5,h\evec{k}}^{\rm cQPA} = \sum_{s={\pm}} \left[
        -\frac{sh|\evec{k}|}{\omega_{\evec{k}}} \, f_{h\evec{k}}^{ms} + \left(
            \frac{h|\evec{k}|m_{\mathrm{R}}}{\omega_{\evec{k}}^2} + \frac{\im sm_{\mathrm{I}}}{\omega_{\evec{k}}}
        \right) f_{h\evec{k}}^{cs}
    \right] \text{.}
\end{equation}
%

%
\subsection{Collisionless case}
\label{sec:collisionless}
%
We first point out that currents computed with the exact Wightman function fully agree with the cQPA currents in the collisionless limit. This may look surprising, because cQPA relies on a spectral ansatz derived to lowest order in gradients. Yet, at the integrated level the collisionless cQPA is in fact \emph{exact} and cQPA shell functions are in one-to-one correspondence with the \emph{local limit} of the correlation functions~\cite{Fidler:2011yq}, and the correspondence is not affected by the introduction of a damping term. This can be illustrated explicitly \eg with equations~\eqref{eq:DampedWightmanFunctionExample} and~\eqref{eq:cQPA-correlator-free-particle}: integrating equation~\eqref{eq:DampedWightmanFunctionExample} over $k_0$ gives
\begin{align}
    \int\frac{\mathrm{d}k_0}{2\uppi}\,\overbar S_{h\evec{k},\Gamma}^<(k_0,t)
    &= \int \frac{\mathrm{d}k_0}{2\uppi}\,\int \mathrm{d}r_0\,\e^{\im k_0r_0 - \Gamma_{h\evec{k}}|r_0|}
        \overbar S_{h\evec{k}}^<(t + \sfrac{r_0}{2}, t - \sfrac{r_0}{2})
    \notag \\
    &= \overbar S^<_{h\evec{k}}(t,t)
    \, \overset{\text{cQPA}}{\longrightarrow} \, \sum_{\pm} \left[
        P^{m\pm}_{h\evec{k}}f^{m\pm}_{h\evec{k}}
      + P^{c\pm}_{h\evec{k}}f^{c\pm}_{h\evec{k}}
   \right] \text{,}
\end{align}
where in the last line we used the cQPA-ansatz~\eqref{eq:cQPA-correlator-free-particle}. Thus, \emph{the essential feature of the cQPA is} not the expansion in gradients or the ensuing spectral approximation, but \emph{the assumption that non-local degrees of freedom are not dynamical}. In particular this result shows that cQPA retains the full quantum information relative to the average time coordinate $t$.

Finally, let us stress the delicate role the decay width $\Gamma$ plays in the emergence of the cQPA-scheme. On one hand, we have seen that if $\Gamma$ was vanishing, non-local temporal correlations would dominate the correlation function; the quality of the local approximation then crucially depends on a non-zero damping. Yet, the spectral limit formally corresponds to taking $\Gamma\rightarrow 0$. That is, $\Gamma$ must be large enough to ensure that non-local correlations can be neglected, and yet small enough so that a spectral quasiparticle picture is valid. Fortunately this is typically the case. We shall elaborate more on these issues in a forthcoming publication~\cite{KKJ_inpreparation}.

%
\subsection{Semiclassical approximation}
\label{sec:semiclassical}
%

While the cQPA is designed to capture the local quantum effects in a generic evolving background, a different method exists for systems in slowly varying backgrounds. The \emph{semiclassical approximation} was introduced in refs.~\cite{Cline:2000nw,Cline:2001rk,Kainulainen:2001cn,Kainulainen:2002th} for systems with spatial inhomogeneities, and the details for temporally varying systems can be found in ref.~\cite{Prokopec:2003pj}. The semiclassical approximation is also local, but in contrast to cQPA, one applies the gradient expansion directly to the \emph{unintegrated} equations of motion, eliminating off-diagonal chiral degrees of freedom. This leads to a loss of information in comparison to cQPA.

We do not get into the details of the derivation, but merely quote the results relevant for our purposes. The Wightman function is decomposed into a helicity block-diagonal form
\begin{equation}
\label{eq:block}
2\im\gamma^0 S^<_{h\evec{k}}(k_0,t) = \sigma^a g_{ah\evec{k}}(k_0, t) \otimes P^{\scriptscriptstyle (2)}_{h\evec{k}} \text{,}
\end{equation}
where $a \in \{0,1,2,3\}$, $\sigma^0 \equiv \idmat$, $\sigma^i$ are the Pauli matrices, and $g_{ah}$ are the unknown coefficient functions to be solved.
The main outcome of the semiclassical formalism is that, when considered to the first order in the gradients of a time-dependent mass $m = |m|\e^{\im\theta}$, the axial part of the helicity correlation function $g_{3h\evec{k}}$ is found to be living on a shifted energy shell: $g_{3h\evec{k}} \sim \delta\left(k_0^2-\omega_{3h\evec{k}}^2\right)$, with
\begin{equation}
\omega_{3h\evec{k}} \equiv \omega_{\evec{k}}(t) + h\frac{|m|^2\partial_t\theta(t)}{2|\evec{k}|\omega_{\evec{k}}(t)}
\text{.}
\end{equation}
The shift has an opposite sign for particles with opposite helicities, and it obviously vanishes for translationally invariant systems.\footnote{For problems with a spatially varying mass a similar shift occurs for the zeroth component $g_{0}$, and is proportional to the spin of the particle~\cite{Kainulainen:2001cn, Kainulainen:2002th}.}

Defining the integrated phase space densities
\begin{equation}
\label{eq:intdens}
f_{ah\evec{k}}(t) \equiv \int \frac{\mathrm{d}k_0}{2\uppi} \, g_{ah\evec{k}}(k_0,t)
\end{equation}
one finds the following collisionless equation of motion for the axial density $f_{3h\evec{k}}$~\cite{Prokopec:2003pj}:
\begin{equation}
\label{eq:f3h}
\left[\omega_{3h\evec{k}}\partial_t + F_{h\evec{k}}\partial_{k_0}\right]f_{3h\evec{k}} = 0
\text{,}
\end{equation}
where $F_{h\evec{k}}$ is the \emph{semiclassical force}
\begin{equation}
F_{h\evec{k}} = \frac{\partial_t|m|^2}{2\omega_{3h\evec{k}}} + h\frac{\partial_t(|m|^2\partial_t\theta)}{2|\evec{k}|\omega_{\evec{k}}}\text{.}
\end{equation}
This process of going from quantum equations (cQPA) to the semiclassical force is analogous to going from the Schrödinger equation to a spin-dependent force when calculating an electron's movement in a magnetic field (the Stern--Gerlach experiment). Noticing that $F_{h\evec{k}} = \partial_t\omega_{3h\evec{k}}$, one can see that the collisionless equation~\eqref{eq:f3h} is solved by
\begin{equation}
\label{eq:semif}
    f_{3h\evec{k}}^{\rm sc}(t)
    = \frac{\omega_{-}f_{3h\evec{k}}^-}{\omega_{3h\evec{k}}(t)}
    \text{,}
\end{equation}
where $f_{3h\evec{k}}^- \equiv f_{3h\evec{k}}(t \rightarrow -\infty)$ is determined by the desired initial conditions. These formulae are valid for an arbitrary form of the mass function. Note that the definition of the phase space function $f_{3h\evec{k}}$ exactly coincides with our definition of the current $j_{5,h\evec{k}}$ in equations~\cref{eq:current,eq:axialDensity}.
%
%
\subsection{Range of validity of the different formalisms}
\label{sec:rangeofvalidity}
%

Let us now compare the axial quantum currents to their semiclassical approximation in different kinematical regions. We use the initial conditions described in section~\ref{sec:phase}, which correspond to choosing $f_{3h\evec{k}}^- = h|\evec{k}|/\omega_{-}$ in equation~\eqref{eq:semif}. In cQPA the equivalent initial configuration for $S^<$ is \(f_{h\evec{k}}^{m-}(-\infty) = 1\) with other shell functions vanishing. In this case the semiclassical approximation gives the following form for the helicity-summed axial density of our kink-mass system:
\begin{equation}
\label{eq:semikink}
j^{\rm sc}_{5,\evec{k}}(t) = \sum_h f_{3h\evec{k}}^{\rm sc}(t) = -\frac{m_{\mathrm{I}}m_{2\mathrm{R}}}{\tau_{\rm w}\,\omega_{\evec{k}}^{3}(t)\cosh^2(\nicefrac{t}{\tau_{\rm w}})}
\text{.}
\end{equation}

\begin{figure}[t!]
 \hskip0.7truecm
  \includegraphics[width=0.85
  \textwidth]{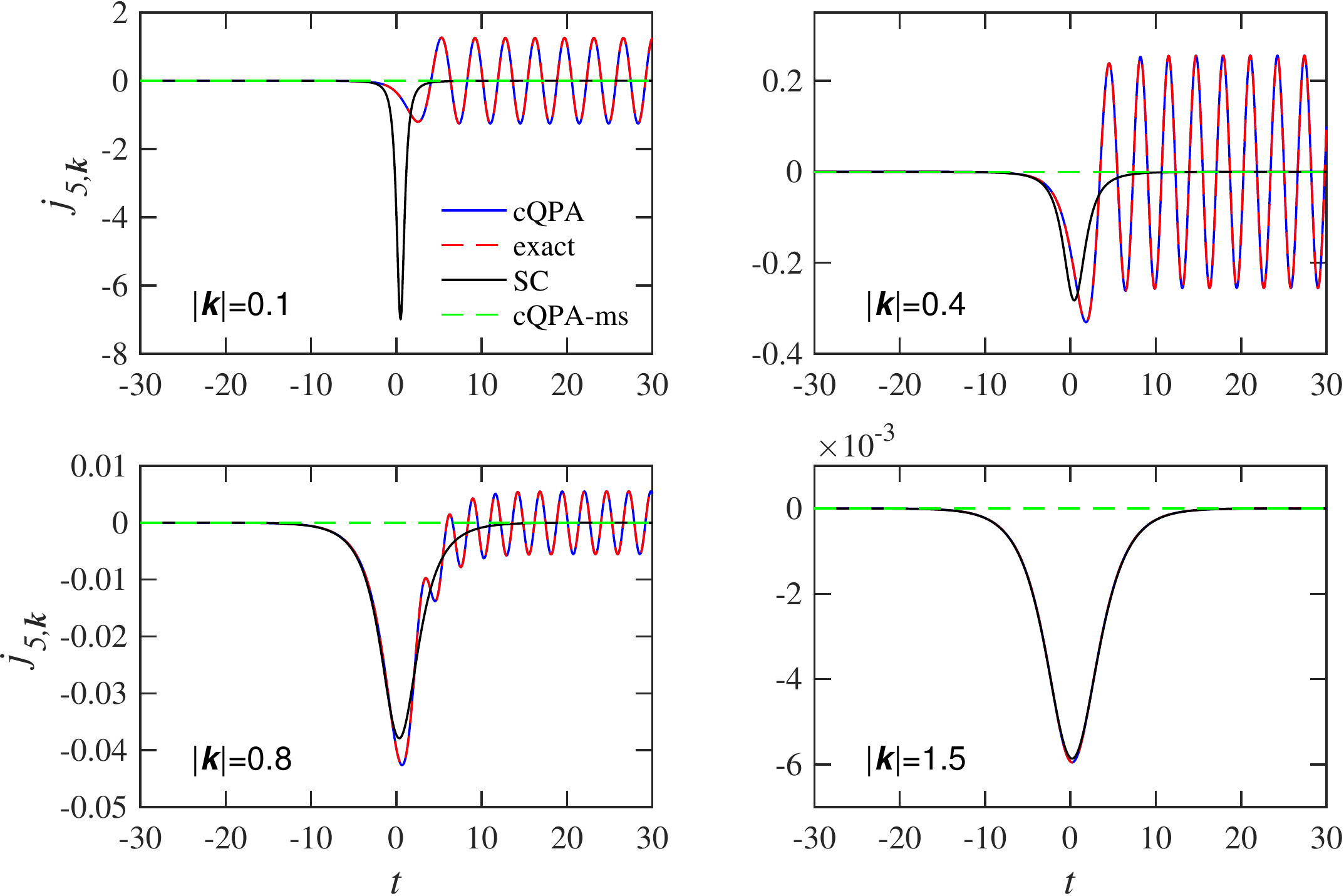}
  \caption{The helicity summed axial charge density $j_{5,\evec{k}}$ from the exact solutions (red dashed line), and from the semiclassical approximation (black line). Blue solid line (exactly matching the red dashed line) is the full cQPA solution and the green line is cQPA solution restricted to the mass shells.
  In each figure we have $m_{1\mathrm{R}}=0.1$, $m_{2\mathrm{R}}=1$, $m_{\mathrm{I}}=0.1$, $\tau_{\rm w}=5$ and $\Gamma=0.2$, while $|\evec{k}| = 0.1,\, 0.4,\, 0.8$ and $1.5$ in different panels as indicated.}
\label{fig:f7}
\end{figure}

In figure~\ref{fig:f7} we show the helicity summed axial density $j_{5,\evec{k}} \equiv \sum_h j_{5,h\evec{k}}$ as a function of time for a few representative values for $|\evec{k}|$, computed from the semiclassical equation~\eqref{eq:semikink}, using our exact solutions with equation~\eqref{eq:current} and using the cQPA methods via equation~\eqref{eq:cqpatosc}. As explained above, the full cQPA-currents coincide with the exact currents in the collisionless limit. In this case the cQPA-current is \emph{pure coherence}, since the cQPA-solution restricted to mass shells (green dashed lines) gives a vanishing axial current.

%
%
%
%

The general comparison to the semiclassical approximation is as expected: prominent oscillations appearing in the exact solutions  for small $|\evec{k}|$ are absent in the semiclassical solution. This is as it should be, since quantum coherence effects are included in the semiclassical formalism only in an average sense. However, the oscillations turn off quickly for large $|\evec{k}|$, such that already for $|\evec{k}| = 1.5$ the semiclassical and quantum currents are practically identical. Moreover, the semiclassical current captures the \emph{average} of the exact solution very well for $|\evec{k}| = 0.8$ and reasonably well even for $|\evec{k}| = 0.4$. The broad range of validity of the semiclassical approximation is slightly surprising. On general grounds one would assume it to work when at least one wavelength fits to the wall width, corresponding to $\frac{2\uppi}{|\evec{k}|} < \tau_{\rm w}$. However, our results suggest that it works quite well even when the wall width is but a fraction of the wave length of the mode.

\begin{figure}[t!]
    \hskip0.7truecm
    \includegraphics[width=0.85\textwidth]{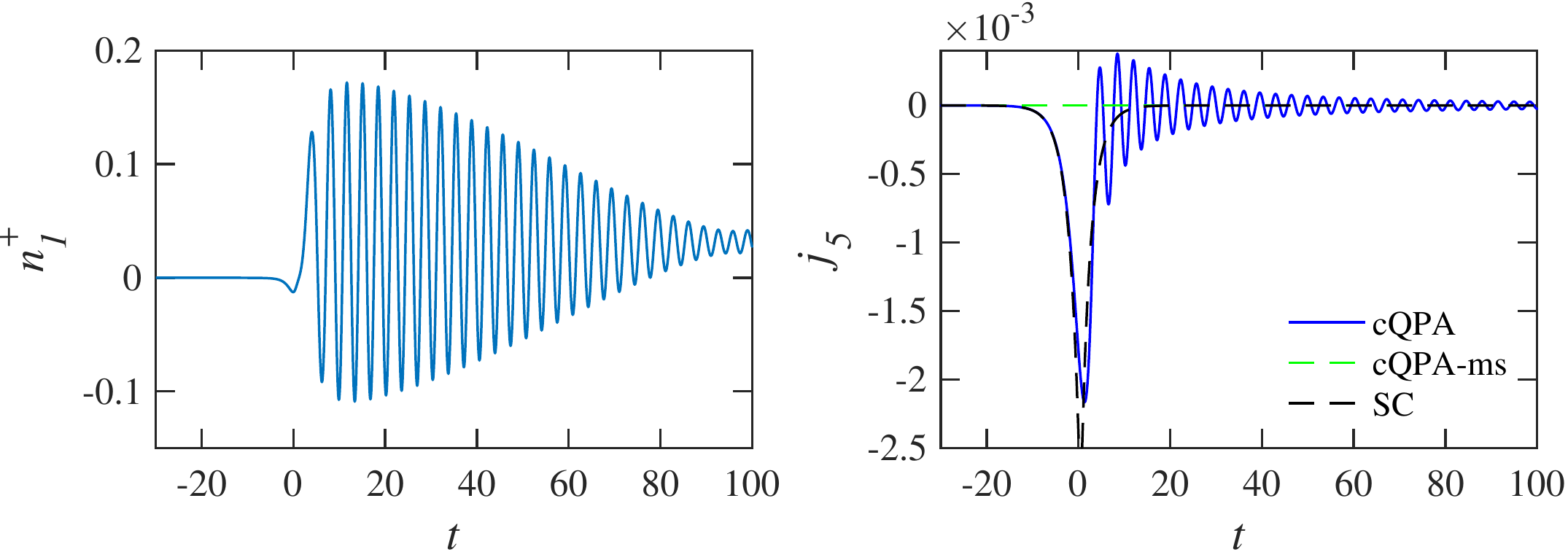}
    \caption{Shown is the integrated number density $n^{+}_1$ of positive helicity particles (left graph) and the integrated axial charge density $j_{5}$ (right graph) for a vacuum initial condition in the non-interacting case. We used the same set of mass parameters as in figure~\ref{fig:f7}.}
    \label{fig:f8}
\end{figure}

The validity of the semiclassical approximation is even more pronounced when one considers the integrated current
\begin{equation}
j_5(t) \equiv \frac{1}{2\uppi^2}\int {\rm d}|\evec{k}|\,\evec{k}^2 \sum_h j_{5,h\evec{k}}(t)\text{.}
\end{equation}
In the right panel of figure~\ref{fig:f8} we show the result of the calculation of $j_5(t)$ for the same set of parameters as considered in figure~\ref{fig:f7}. Apart from the oscillations right after the mass change, the semiclassical solution follows the full solution quite well. In the left panel we show the behaviour of the integrated number density $n^{+}_1$ of positive helicity particles. (The individual number densities are defined below in section~\ref{sec:cQPA_with_collisions}.) Indeed, oscillations tend to be much larger in the individual components, but they mostly cancel out at the level of currents.

%
%
%
%

Our results in the non-interacting case are qualitatively similar to those of ref.~\cite{Prokopec:2013ax}; the semiclassical approximation captures the mean trend of the currents quite well. However,
while ref.~\cite{Prokopec:2013ax} emphasized the fact that the semiclassical approximation misses the late time oscillations, we do not think that this is necessarily a significant problem. First, we see that the oscillations damp quite quickly. Second, a typical application of a calculation presented here would be to compute the particle-antiparticle asymmetry arising from the transition. The axial current would then be closely related to the source of the asymmetry. In such a case the effect of oscillations around the mean would tend to cancel out, leaving a mean effect that could be well captured by the semiclassical result.

Let us emphasize that the cQPA result for the current indeed contains and generalises the semiclassical result. This is so despite the fact that the cQPA-dispersion relation was derived formally to lower order in gradients than the semiclassical one. The reason for this apparently contradicting result was already emphasized in the beginning of this section: at the integrated level the non-interacting cQPA is in fact exact. Similarly then, the interacting cQPA-equations~\eqref{eq:cqpa_singleflav1} constitute a generalisation of the interacting semiclassical Boltzmann theory to the fully quantum case. We now turn to study such interacting systems in the context of cQPA. This requires that we define explicitly the collision terms in equations~\eqref{eq:cqpa_singleflav1}.

%
\section{cQPA with collisions}
\label{sec:cQPA_with_collisions}
%

Let us now assume that the self-energy satisfies the KMS-relation $\Sigma^> = \e^{\beta k_0}\Sigma^<$. This is perhaps the most often recurring application, so we write down the full single flavour interacting cQPA-equations~\eqref{eq:cqpa_singleflav1} explicitly for this case. After some algebra we find:
\begin{subequations}
\label{eq:cqpa_singleflav}
\begin{align}
\label{eq:cqpa_singleflav_a}
\partial_t n_{h\evec{k}}^{\pm} ={}& \frac{1}{2}\sum_s \dot{\Phi}_{h\evec{k}}^s \, f_{h\evec{k}}^{cs} - \sum_s \left[\bigl(n_{h\evec{k}}^s - n_{\mathrm{eq}}^s\bigr) T_{mm}^{hs\pm} + f_{h\evec{k}}^{cs} \, T_{cm}^{hs\pm}\right] \text{,} \\
\label{eq:cqpa_singleflav_b}
\partial_t f_{h\evec{k}}^{c\pm} ={}& \mp 2\im\omega_{\evec{k}} f_{h\evec{k}}^{c\pm} + \xi_{\evec{k}} \dot{\Phi}_{h\evec{k}}^{\mp} \left[\frac{m_{\mathrm{R}}}{\omega_{\evec{k}}} f_{h\evec{k}}^{c\pm} + \frac{1}{2}\bigl(1 - n_{h\evec{k}}^+ - n_{h\evec{k}}^-\bigr) \right] \\ \notag
&{}- \xi_{\evec{k}}\sum_s  \left[\bigl(n_{h\evec{k}}^s - n_{\mathrm{eq}}^s\bigr) T_{mc}^{hs\pm} + f_{h\evec{k}}^{cs} \, T_{cc}^{hs\pm}\right]
\text{,}
\end{align}
\end{subequations}
where $\dot{\Phi}_{h\evec{k}}^{\pm}$ and $\xi_{\evec{k}}$ were defined in equation~\eqref{eq:phidots} and we replaced the mass shell functions by the number densities $n^+_{h\evec{k}} \equiv f^{m+}_{h\evec{k}}$ and $n^-_{h\evec{k}}\equiv 1-f^{m-}_{h\evec{k}}$ (these are the usual 1-particle Boltzmann distribution functions) and $n_{\rm eq}^s \equiv f_{\rm eq}({+}\omega_{\evec{k}})$. Finally, the $T^{hs\pm}_{ab}$-functions encode the collision terms for generic thermal interactions. In the spatially homogeneous and isotropic system the most general form of the self-energy function can be expanded as
\begin {equation}
    \Sigma^{\mathcal{A}}_{\evec{k}}(k_0,t) \equiv \sum_i c^{\mathcal{A}}_i(k,t) \sigma_i(k) \text{.}
    \label{eq:general-SigmaA}
\end{equation}
Here $\sigma_i(k)$ are the Dirac structures given in the leftmost column of table~\ref{table:coll} and $c^{\mathcal{A}}_i(k,t)$ are some four-momentum- and possibly time-dependent functions.\footnote{Note that the last four rows in table~\ref{table:coll} contain redundant information. For example, using the fact that $\slashed k P_{h{\evec{k}}} = (k_0 \gamma^0 - h|\evec{k}|\gamma^0\gamma^5) P_{h\evec{k}}$, one finds that $(\mathcal{T}_{\slashed k})_{ab}^{hss'} = \omega_{\evec{k}} (\mathcal{T}_{\sgn(k_0)\gamma^0})_{ab}^{hss'} - h|\evec{k}|(\mathcal{T}_{\gamma^0\gamma^5})_{ab}^{hss'}$. It is easy to check that this relation is satisfied by the entries of table~\ref{table:coll}. Similarly $\frac{1}{2}[\gamma^0,\slashed  k]P_{h{\evec{k}}} = -h|\evec{k}|\gamma^5P_{h{\evec{k}}}$, which implies that the last two rows are just $-h|\evec{k}|$ times the first two lines in reverse order. However, rather than being minimalistic, we give a complete list of the possible structures.} Interaction terms corresponding to equation~\eqref{eq:general-SigmaA} are given by
\begin{equation}
\begin{aligned}
    T^{hss'}_{mm}(|\evec{k}|,t) &= \sum_i c^{\mathcal{A}}_i(s)({\mathcal{T}}_i)^{hss'}_{mm}(|\evec{k}|) \text{,} \\
    T^{hss'}_{cm}(|\evec{k}|,t) &= \sum_i\biggl[
        \frac{c^{\mathcal{A}}_i(s)+c^{\mathcal{A}}_i(-s)}{2}
        - ss'\frac{c^{\mathcal{A}}_i(s)-c^{\mathcal{A}}_i(-s)}{2}
    \biggr](\mathcal{T}_i)^{hss'}_{cm}(|\evec{k}|) \text{,} \\
    T^{hss'}_{mc}(|\evec{k}|,t) &= \sum_i c^{\mathcal{A}}_i(s) (\mathcal{T}_i)^{hss'}_{mc}(|\evec{k}|) \text{,} \\
    T^{hss'}_{cc}(|\evec{k}|,t) &= \sum_i \frac{c^{\mathcal{A}}_i(s)-c^{\mathcal{A}}_i(-s)}{2}(\mathcal{T}_i)^{hss'}_{cc}(|\evec{k}|) \text{,}
\end{aligned}
\label{eq:cqpa-collision-terms}
\end{equation}
where $c^{\mathcal{A}}_i(s)\equiv c^{\mathcal{A}}_i(s\omega_{\evec{k}},|\evec{k}|,t)$ and the functions $(\mathcal{T}_i)^{hss'}_{ab}$ can be read from table~\ref{table:coll}, where we further defined
\begin{align}
    A_{h\evec{k}}^s &\equiv h \frac{|\evec{k}| m_{\rm R}}{\omega_{\evec{k}}^2}
        + \im s \frac{m_{\rm I}}{\omega_{\evec{k}}} \text{,} \\
    B_{h\evec{k}}^s &\equiv s h \frac{|\evec{k}|}{\omega_{\evec{k}}}
        + \im \frac{m_{\rm R}m_{\rm I}}{\omega_{\evec{k}}^2} \text{.}
\end{align}
%

%
%
\begin{table}[t!]
\begin{center}
\begin{tabular}{lr@{}lr@{}lr@{}lr@{}l}
    \toprule
    $\sigma_i$ & & $(\mathcal{T}_i)_{mm}^{hss'}$ & & $(\mathcal{T}_i)_{cm}^{hss'}$ & & $(\mathcal{T}_i)_{mc}^{hss'}$ & & $(\mathcal{T}_i)_{cc}^{hss'}$ \\
    \midrule
    $\idmat$ & & $2s\delta_{ss'}\frac{m_{\rm R}}{\omega_{\evec{k}}}$ & & $s'/\xi_\evec{k}$ & & $s/\xi_\evec{k}$ & & $2s\delta_{ss'}\frac{m_{\rm R}}{\omega_{\evec{k}}}/\xi_{\evec{k}}$ \\[0.7em]
    $\gamma^5$ & $-$ & $2\im s\delta_{ss'}\frac{m_{\mathrm{I}}}{\omega_{\evec{k}}}$ & & $s'B^s_{h\evec{k}}$ & & $sB^{-s'}_{h\evec{k}}$ & $-$ & $2\im s\delta_{ss'}\frac{m_{\rm I}}{\omega_{\evec{k}}}/\xi_{\evec{k}}$ \\[0.7em]
    $\sgn(k_0)\gamma^0$ & & $2s\delta_{ss'}$ & & $0$ & & $0$ & & $2s\delta_{ss'}/\xi_\evec{k}$ \\[0.7em]
    $\gamma^0\gamma^5$ & & $2s\delta_{ss'}h\frac{|\evec{k}|}{\omega_{\evec{k}}}$ & $-$ & $s'A_{h\evec{k}}^s$ & $-$ & $sA_{h\evec{k}}^{-s'}$ & & $2s\delta_{ss'}h\frac{|\evec{k}|}{\omega_{\evec{k}}}/\xi_{\evec{k}}$ \\[0.7em]
    $\slashed{k}$ & & $2s\delta_{ss'}\frac{|m|^2}{\omega_{\evec{k}}}$ & & $s'h|\evec{k}|A_{h\evec{k}}^s$ & & $sh|\evec{k}|A_{h\evec{k}}^{-s'}$ & & $2s\delta_{ss'}\frac{|m|^2}{\omega_\evec{k}}/\xi_\evec{k}$ \\[0.7em]
    $\slashed{k}\gamma^5$ & & $0$ & & $\omega_{\evec{k}}A_{h\evec{k}}^s$ & $-$ & $\omega_{\evec{k}}A_{h\evec{k}}^{-s'}$ & & $0$ \\[0.7em]
    $\sfrac{1}{2}[\gamma^0,\slashed k]$ & & $2\im s\delta_{ss'}h|\evec{k}|\frac{m_{\mathrm{I}}}{\omega_{\evec{k}}}$ &$-$& $s'h|\evec{k}|B^s_{h\evec{k}}$ &$-$& $sh|\evec{k}|B^{-s'}_{h\evec{k}}$ & & $2\im s\delta_{ss'}h|\evec{k}|\frac{m_{\rm I}}{\omega_{\evec{k}}}/\xi_{\evec{k}}$ \\[0.7em]
    $\sfrac{1}{2}[\gamma^0,\slashed k]\gamma^5$ &$-$& $2s\delta_{ss'}h|\evec{k}|\frac{m_{\rm R}}{\omega_{\evec{k}}}$ &$-$& $s'h|\evec{k}|/\xi_\evec{k}$ &$-$& $sh|\evec{k}|/\xi_\evec{k}$ &$-$& $2s\delta_{ss'}h|\evec{k}|\frac{m_{\rm R}}{\omega_{\evec{k}}}/\xi_{\evec{k}}$ \\
    \bottomrule
\end{tabular}
\end{center}
\caption{Collision term coefficients for different self-energy components $\sigma_i(k)$ of $\Sigma_{\evec{k}}$.}
\label{table:coll}
\end{table}
%
%

The collision terms of equations~\eqref{eq:cqpa-collision-terms} together with table~\ref{table:coll} allow for completely general coefficient functions $c_i(k,t)$ of the self-energy~\eqref{eq:general-SigmaA}. However, in thermal equilibrium the functions $c_i(k,t)$ are typically either even or odd functions of $k_0$.
As an example, we consider a thermal self-energy with a chiral interaction given by
\begin{equation}
    \Sigma_{\evec{k}}^\mathcal{A}(k_0) = (a\slashed k + b\slashed u)P_{\rm L} \text{,}
\label{eq:samplesigma}
\end{equation}
where \(u_{\mu}\) is the fluid four-velocity. We further assume that, in the rest frame of the thermal plasma where $\slashed u\to\gamma^0$, the coefficient $a=a(k_0,|\evec{k}|)$ is an odd and $b=b(k_0,|\evec{k}|)$ an even function of $k_0$. Using table~\ref{table:coll}, we then get the following collision terms for equations~\eqref{eq:cqpa_singleflav}:
\begin{equation}
\begin{aligned}
    T^{hss'}_{mm}(|\evec{k}|,t) &= \textstyle
        \Bigl[
            \frac{|m|^2}{\omega_{\evec{k}}}a_{\evec{k}} + \left(1 - sh\frac{|\evec{k}|}{\omega_{\evec{k}}}\right)b_{\evec{k}}
        \Bigr]\delta_{ss'} \text{,} \\
    T^{hss'}_{cm}(|\evec{k}|,t) &= \textstyle
        \frac{s'}{2}\Bigl[
            (\omega_{\evec{k}} - s'h|\evec{k}|)a_{\evec{k}} + b_{\evec{k}}
        \Bigr] A^s_{h\evec{k}} \text{,} \\
    T^{hss'}_{mc}(|\evec{k}|,t) &= \textstyle
        \frac{s}{2}\Bigl[
            (\omega_{\evec{k}} + sh|\evec{k}|)a_{\evec{k}} + b_{\evec{k}}
        \Bigr]A^{-s'}_{h\evec{k}} \text{,} \\
    T^{hss'}_{cc}(|\evec{k}|,t) &= \textstyle
        \frac{1}{\xi_{\evec{k}}}\Bigl[
            \frac{|m|^2}{\omega_{\evec{k}}}a_{\evec{k}} + b_{\evec{k}}
        \Bigr]\delta_{ss'} \text{.}
\end{aligned}
\end{equation}
Here $a_{\evec{k}}\equiv a(\omega_{\evec{k}},|\evec{k}|)$, $b_{\evec{k}}\equiv b(\omega_{\evec{k}},|\evec{k}|)$ and we used the parity properties $a(s\omega_{\evec{k}},|\evec{k}|)=sa_{\evec{k}}$ and $b(s\omega_{\evec{k}},|\evec{k}|)=b_{\evec{k}}$. Also, given that $a_{\evec{k}},b_{\evec{k}}>0$, note how $T^{hss'}_{mm}$ and $T^{hss'}_{cc}$ are always positive.

Let us finally point out that it is easy to generalise equations~\eqref{eq:cqpa_singleflav} to the case with a non-thermal self-energy that does \emph{not} obey the KMS-relation. One just needs to replace the two terms involving the equilibrium distribution function $n^s_{\mathrm{eq}}$ as follows:
\begin{equation}
(n^s_{h\evec{k}} - n^s_{\mathrm{eq}}) T^{hss'}_{ma}(|\evec{k}|,t)
\rightarrow
\sum_i s \Bigl(f^{ms}_{h\evec{k}} c^{\mathcal{A}}_i(s) - \frac{1}{2}c^{<}_i(s)\Bigr) ({\mathcal{T}}_i)^{hss'}_{ma}(|\evec{k}|,t)
\end{equation}
for $a = m,c$, where we defined  $\im\Sigma^<_{\evec{k}}(k_0,t) \equiv \sum_i c_i^<(k,t) \sigma_i(k)$. We remind, however, that evaluating the self-energy diagrams involving coherent propagators as internal lines requires special techniques developed in refs.~\cite{Herranen:2010mh,Fidler:2011yq}.

%
\subsection{A numerical example}
%

\begin{figure}[t!]
 \hskip0.7truecm
  \includegraphics[width=0.85
  \textwidth]{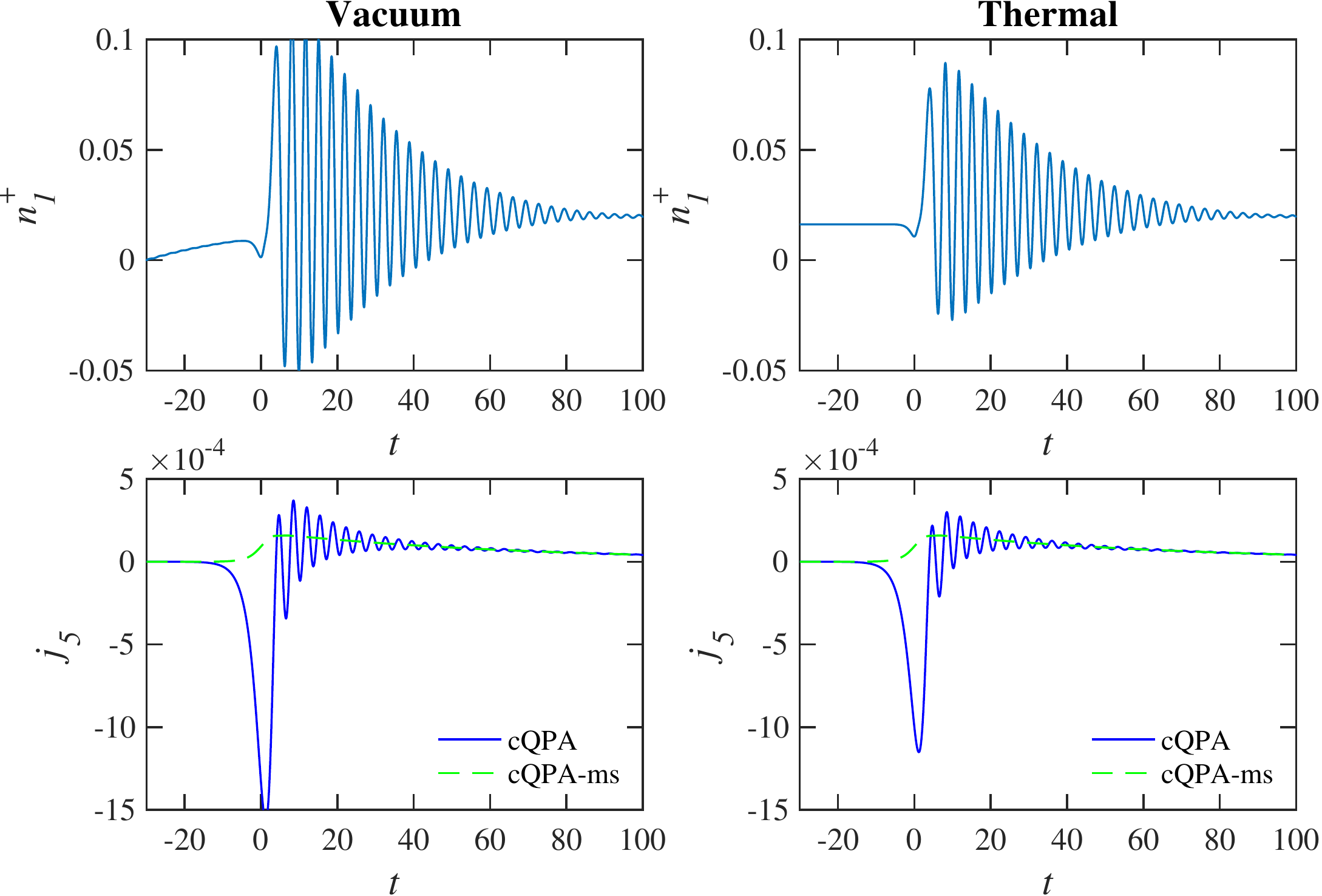}
  \caption{Shown is the integrated number density $n^+_1$ of positive helicity particles (the upper panels) and the integrated axial charge density $j_{5}$ (the lower panels) in interacting cQPA. The left panels correspond to the vacuum initial condition and the right panels to the thermal initial condition with $T=1$. We used the same set of mass parameters as in figure~\ref{fig:f7}.}
\label{fig:f9}
\end{figure}

In figure~\ref{fig:f9} we show a result of a model calculation with a non-vanishing interaction rate using a self-energy of the form~\eqref{eq:samplesigma} with $a_\evec{k}= 0.03$ and $b_\evec{k}=0$. The left panels, where we imposed the vacuum initial conditions $n^\pm_{h\evec{k}}=f^{c\pm}_{h\evec{k}}=0$, correspond to the interacting version of the case studied in figure~\ref{fig:f8}. Initially, the particle number approaches smoothly the thermal value. At the onset of the transition it again starts oscillating, but the amplitude is strongly damped in comparison to the non-interacting case. In the right panels we show the analogous calculation with equilibrium initial conditions $n^\pm_{h\evec{k}} = n_{\rm eq}^\pm$ with $T = 1$ in the units we are working with and $f^{c\pm}_{h\evec{k}}=0$. Now the particle number stays unchanged until the onset of the transition, after which it oscillates approaching asymptotically the same post-transition equilibrium value as in the case with vacuum initial conditions. Pushing the starting point further away from the transition region would make the later evolution indistinguishable in the two cases.

%
%
%
%

The main difference to the non-interacting case is that the left-chiral interaction, in connection with the coherent CP-violating oscillations, creates a temporary non-zero average chiral current after transition. This is due to the fact that the chiral interaction term~\eqref{eq:samplesigma} breaks the helicity symmetry. The average current is well captured at late times by the pure mass shell contribution, shown in green dashed line in figure~\ref{fig:f9}. However at the transition point the main peak is still pure coherence.
While the current eventually equilibrates to zero, the region where it is non-vanishing could act as a seed for example for a particle-antiparticle asymmetry creation in such a transition. 

The calculation we presented here was just a toy model whose sole purpose was to show how to implement the method and display some of the effects of interactions. There are several interesting applications for the formalism that we shall pursue in the future. One avenue is the study of baryogenesis in abrupt spatially homogeneous phase transitions in the early universe, such as the models considered in the context of the cold baryogenesis~\cite{Tranberg:2009de,Tranberg:2012qu,Tranberg:2012jp}. Another application is to study the reheating phase after inflation. It is straightforward to couple equations~\eqref{eq:cqpa_singleflav} with an equation of motion for the inflaton and model the reheating phase including all quantum effects and interactions.  Our formalism, extended to the flavour mixing case~\cite{Fidler:2011yq}, can also be applied to the study of leptogenesis. It is of particular interest to compare our approach with several other transport theory formulations that also employ the closed time path (CTP) methods, such as those presented in refs.~\cite{Beneke:2010wd,Beneke:2010dz,Garbrecht:2011aw,Garbrecht:2018mrp,Dev:2017wwc,Garny:2011hg,Kartavtsev:2015vto,Dev:2014wsa}.

%
\section{Conclusions and outlook}
\label{sec:conclusions}
%

We have studied the phase space structure of a fermionic two-point function with a varying complex mass. We computed the Wightman function of a non-interacting system for a specific mass profile, and demonstrated that its phase space contains, in addition to the usual mass shell solutions, a shell-like structure located at $k_0 = 0$. This zero-momentum shell describes local-in-time quantum coherence between particles and antiparticles and it was discovered earlier in the context of the cQPA-formalism~\cite{Herranen:2008hu,Herranen:2008di,Herranen:2009zi,Herranen:2009xi,Herranen:2010mh,Fidler:2011yq,Herranen:2008hi}. However, our present derivation did not rely on any approximations, but derived the free Wightman function from the exact mode functions of the system.

In addition to the cQPA-solutions we found other, non-local coherence structures in the exact Wightman function. These structures look peculiar, appearing to let the system become aware of the transition before it actually takes place in the local time coordinate, but of course they are just a reflection of the usual quantum non-locality in the phase space picture. We argued that the non-local correlations would dominate the phase space structure in large non-dissipative systems. However, when dissipation is included (modelled here by a damping term coupled to the relative time coordinate), the non-locality gets confined to the neighbourhood of the transition region. These results underline the delicate role of dissipation in the emergence of the local (cQPA) limit, and eventually (in the nearly translationally invariant systems) of the familiar Boltzmann transport theory.

In section~\ref{sec:Wfunction} we introduced a new and particularly useful way to reorganise the gradient expansion in the mixed representation Kadanoff--Baym equations. Then, based on this form, we gave a simple and transparent derivation of the cQPA equations. In section~\ref{sec:cQPA_with_collisions} we completed the analysis by providing explicit collision integrals for generic interaction self-energies. The resulting equations~\eqref{eq:cqpa_singleflav} are one of the main results of this paper: they generalise the Boltzmann transport theory to systems with local coherence between particles and antiparticles. In particular they fully encompass the well known semiclassical effects. Such coherences may be relevant for example for baryogenesis during phase transitions and for particle production at the end of inflation.

We further computed axial phase space densities out of the Wightman functions and compared these to the same quantities obtained from the semiclassical approximation. We found out that the semiclassical methods work reasonably well even in systems where the relevant modes have wavelengths down to a half of the wall width. This is encouraging for baryogenesis studies in very strong electroweak phase transitions, often encountered in the context of models producing large, observable gravitational wave signals~\cite{Dorsch:2016nrg,Vaskonen:2016yiu}.

In this work we only considered a time-dependent mass. A natural follow-up, relevant for the baryogenesis problem, would be to generalise the analysis to a mass depending on one spatial coordinate. Part of this program is straightforward, but some new features emerge as well, such as the tunneling solutions, whose proper description at the phase space level is non-trivial. But there are practical applications of the time-dependent formalism as well, which we shall be pursuing. One is the baryogenesis at a phase transition as discussed in section~\ref{sec:cQPA_with_collisions} and already studied in the context of a simple toy model in ref.~\cite{Herranen:2010mh}. Another immediate goal is to use equations~\eqref{eq:cqpa_singleflav}, coupled to the one-point function of the inflaton, to model accurately the reheating phase at the end of the inflation. Also, we are pursuing a generalisation of the present formalism to the case with mixing fermion fields, in the context of resonant leptogenesis~\cite{JKR_inpreparation}.

\section*{Acknowledgements}

This work was supported by the Academy of Finland grant 318319. HJ was in addition supported by grants from the Väisälä Fund of the Finnish Academy of Science and Letters and OK by a grant from the Magnus Ehrnrooth Foundation. We wish to thank Pyry Rahkila and Werner Porod for many enlightening discussions.

\bibliographystyle{JHEP}
\bibliography{Wightman}

\end{document}